\documentclass[a4paper,12pt]{article}

\usepackage[pdftex]{graphicx}
\usepackage{subfigure}

\newcommand{\be}{\begin{equation}}
\newcommand{\ee}{\end{equation}}

\newcommand{\bea}{\begin{eqnarray}}
\newcommand{\eea}{\end{eqnarray}}

\newcommand{\p}{\partial}

\newcommand{\nn}{\nonumber \\}
\newcommand{\f}{\frac}
\newcommand{\w}{\wedge}
\newcommand{\ra}{\rightarrow}
\begin{document}
\thispagestyle{empty}
\begin{flushright}
{\bf arXiv:1202.nnnn}
\end{flushright}
\begin{center} \noindent \Large \bf
Approximate strange metallic behavior in AdS 
\end{center}

\bigskip\bigskip\bigskip
\vskip 0.5cm
\begin{center}
{ \normalsize \bf   Shesansu Sekhar Pal\footnote{Permanent address: Barchana,  754296, Jajpur, Odisha, India.}}


\vskip 0.5 cm
Center for Quantum Spacetime,
Sogang University, 121-742, Seoul, South Korea
\vskip 0.5 cm
\sf { shesansu${\frame{\shortstack{AT}}}$gmail.com }
\end{center}
\centerline{\bf \small Abstract}

We show for unit dynamical exponent, $z=1$, the appearance of the Fermi liquid and non-Fermi liquid behavior as we tune the charge density and the magnetic field in $3+1$ dimensional field theory using the gauge-gravity duality. There exists an upturn behavior of the resistivity only along the direction perpendicular to the magnetic field. Also, there exists a universal behavior of the resistivity, independent of the dimensionality of the  spacetime, 
in a specific corner of the parameter space, namely, 
 in the large charge density and small magnetic field limit: the longitudinal conductivity goes as $T^{-2/z}$, whereas the Hall conductivity goes as $T^{-4/z}$. It means the Hall coefficient goes as $T^{4/z}$. We compute the diffusion constant from the flow equation of the conductivity.

\newpage
\tableofcontents

\section{Introduction}
\begin{figure}[htb]
\centering
\subfigure[Generic phase diagram]
{\includegraphics[width=3.0 in] {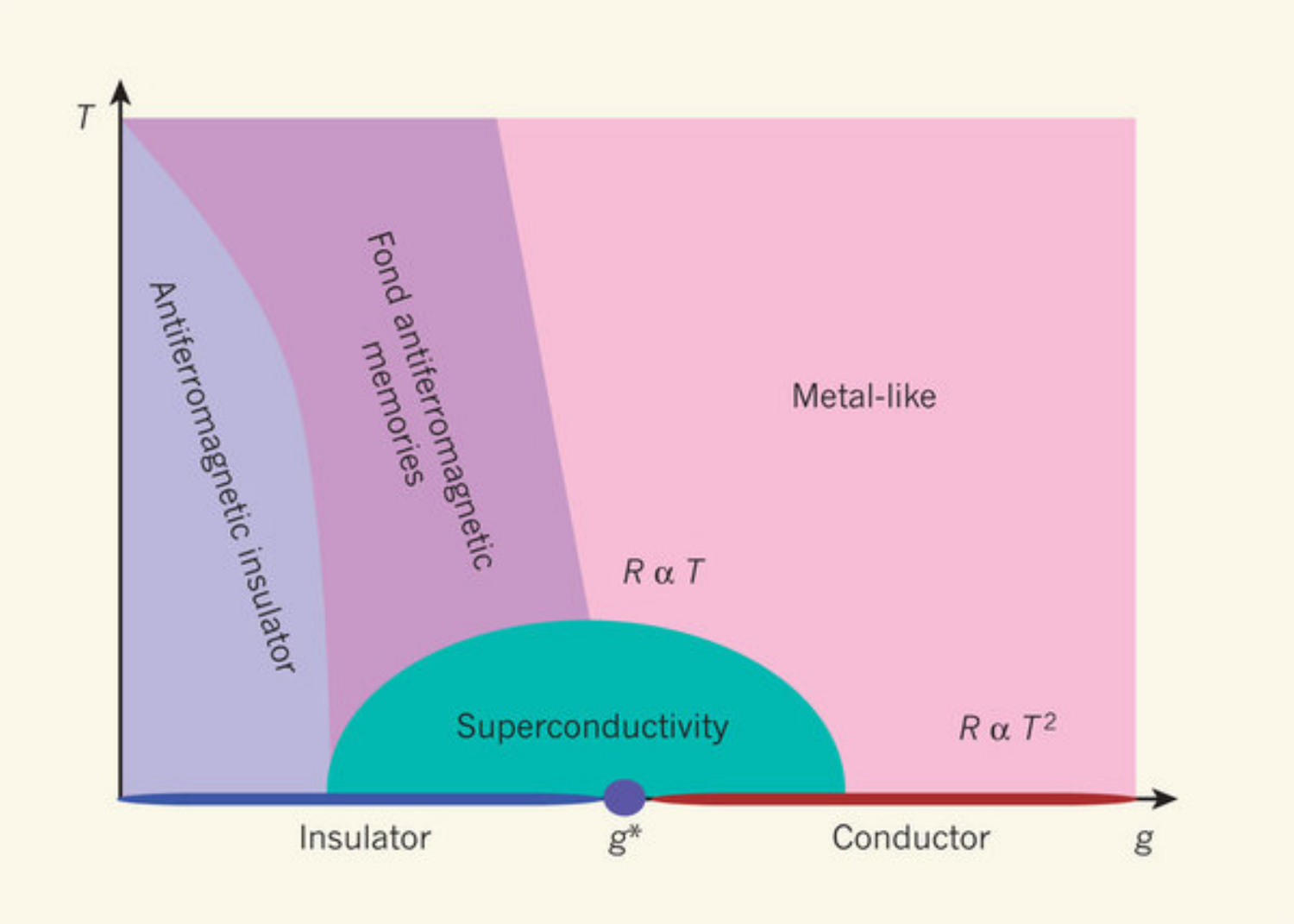}}
\subfigure[Phase diagram with proper scale]
{\includegraphics[width=2.4 in] {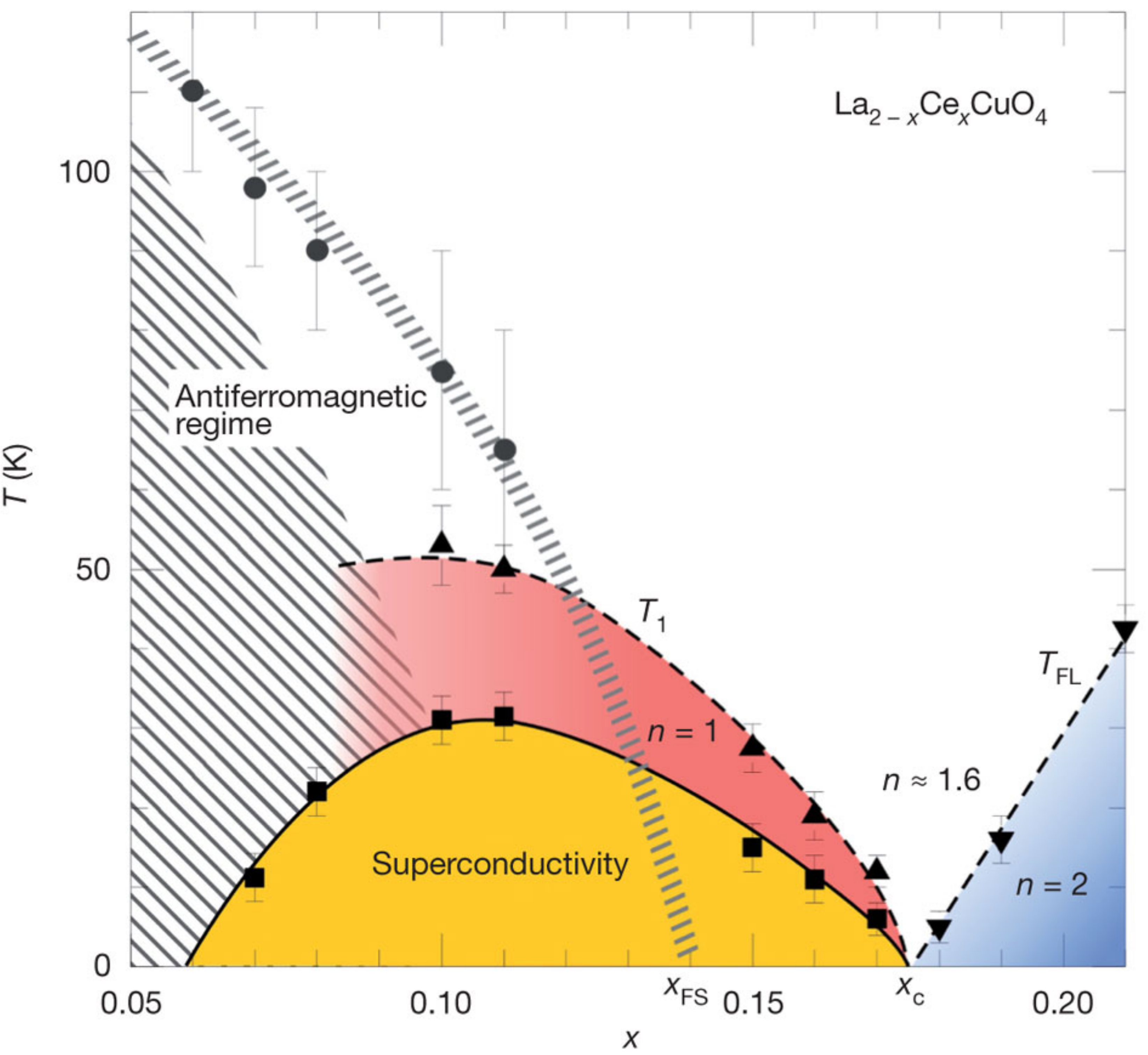}}
\caption{The generic behavior of the phase diagram of the CuO system shown in figure (a) is borrowed from \cite{pmg}. The second figure on the right is borrowed from \cite{kj}. In this figure T stands for the temperature and both $x$ and $g$ denote the doping. The point $g^{\star}$ and $x_c$ denote the quantum critical point.}
\label{fig_1}
\end{figure}

The study of the phase diagram of  superconducting copper oxide system at high temperature has provided some interesting and rich phases. The generic structure of the phase diagram can be summarized in figure (a) of fig(\ref{fig_1}), which is borrowed from \cite{pmg}.
The part of the figure that we are interested in this paper corresponds to the equations written in figure (a). In figure (b), it is indicated with the value of $n$ that appear in the definition of the resistivity,   defined as follows: ${\cal R}={\cal R}_0+A T^n$, where ${\cal R}_0$ is a constant, $A$ can be a function of the charge density and $T$ is the temperature. In what follows, we are  interested only in the particular temperature dependence of the resistivity. It follows,  from fig(\ref{fig_1}), that the Fermi liquid (FL) corresponds to a phase whose resistivity  has the quadratic dependence on the temperature whereas  non-Fermi liquid (NFL) has linear dependence on the temperature.  Moreover, it is suggested in \cite{kj}  there exists a  cross-over from FL phase to the NFL phase and it is absolutely clear from the figure that the FL phase appears at a higher charge density than the NFL phase. In between the NFL and FL phase there occurs a phase whose resistivity goes as $\rho\sim T^{5/3}$.
In what follows, we shall construct a model using the AdS/CFT correspondence \cite{jm} to show the presence of both the FL and the NFL phase  and we can go from one phase to the other by doing appropriate fine tuning  of the parameters like charge density and the magnetic field at low temperature. Moreover, the occurrence of the  FL and NFL phase is possible  only for the  asymptotically AdS spacetime. This is completely different\footnote{Where the authors have shown that the NFL phase is possible only for the Lifshitz type of  spacetime \cite{klm} and \cite{pal}.} from that studied  e.g. in \cite{hpst},\cite{kachru},\cite{kiritsis},\cite{ks}, \cite{ssp},\cite{kkp},\cite{kim}, and \cite{lpp}. The only requirement to see such a behavior of both the FL and the NFL phase  is to work in a $4+1$ dimensional bulk AdS spacetime. There exists another motivation to consider the asymptotically AdS spacetime: in a recent study using the bottom up approach, it is shown that the Lifshitz spacetimes are unstable \cite{cm} and \cite{hw}. So, it suggests to find
other models where we can use the asymptotically AdS spacetimes to construct both the FL and NFL phases.

\begin{figure}[htb]
\centering
\subfigure[Resistivity as a function of temperature]
{\includegraphics[width=3.0 in]
{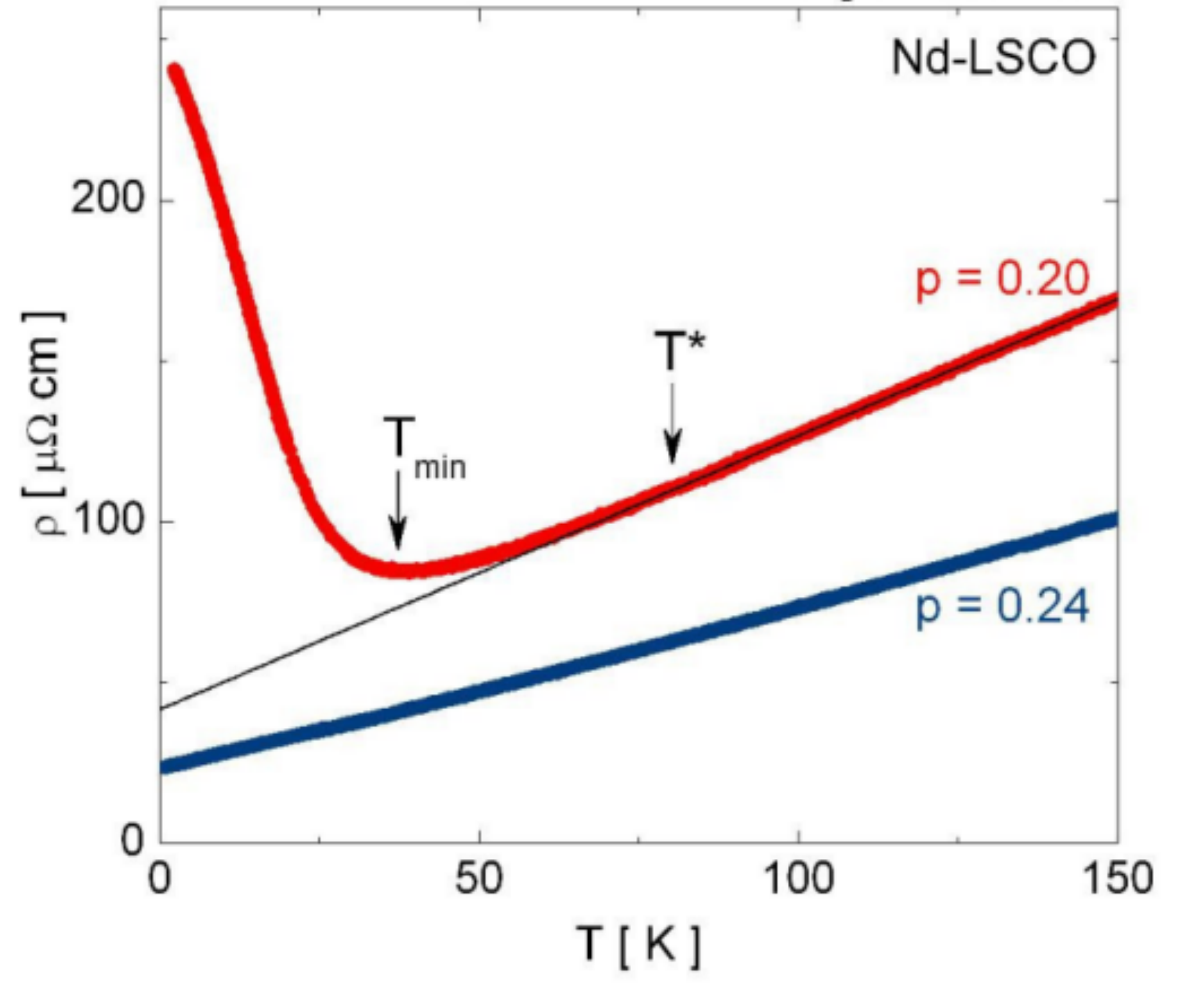}}
\caption{The  behavior of the resistivity as a function of temperature with specific doping $p$ borrowed from \cite{daou}. $T_{min}$ is the temperature below which there occurs a upturn and below $T^{\star}$ the resistivity ceases to behave linear in $T$. }
\label{fig_2}
\end{figure}

In a recent study, Daou {\it et al.} \cite{daou} has found a very interesting behavior of the electrical resistivity as a function of the doping for the NFL phase. For a specific value of the doping, the resistivity shows linear temperature dependence at high temperature and at low temperature it makes an upturn. The upturn behavior of the resistivity   is absent  as one changes the doping. This particular property is shown in  fig(\ref{fig_2}), below.

In this paper,  we study  the finite charge density phases at non-zero temperature  using holography \cite{jm}. It is done by explicitly computing the transport property of these phases, i.e. its dependence on the charge density and temperature etc. In particular, we  calculate the flow equation of the conductivity
in $2+1$ dimensional field theory by considering both the Dirac-Born-Infeld (DBI) and Chern-Simon (CS) action with constant axion. There follows some interesting outcome, i.e. the presence of the axion makes the derivative of the $xy$ component of the resistivity with  respect to the magnetic field to go either as $\f{\p R_{xy}}{\p B}\sim T^{-4/z}$ or as $T^{-8/z}$ for large charge density and small magnetic field limit depending on whether $B^2< T^{4/z}$ or $B^2> T^{4/z}$, whereas   $\f{\p R_{xx}}{\p B}\sim T^{-2/z}$. In $3+1$ dimensional field theory, (1) we show that it is possible to find the resistivity of either the Fermi liquid or the non-Fermi liquid state by tuning the parameters like charge density and magnetic field. (2)  The conductivity along the direction of  the magnetic field does not show any upturn behavior whereas along the perpendicular direction we do see the upturn behavior. Even though we see the upturn behavior for a particular component of the resistivity but it is {\it a priori} not clear what plays the role of the doping parameter $p$ as shown in fig(\ref{fig_2}).   (3) In the $T\ra 0$ limit, the resistivity shows an interesting behavior, namely, the resistivity along the direction of the magnetic field vanishes whereas in the perpendicular direction  diverges.
(4) Then we make a comparative study of the conductivity in $2+1$
dimensional field theory and in $3+1$ dimensional field theory with rotational symmetry, which is summarized below in table 1.

\begin{tabular}{  |l | l |l|}
    \hline
 Field theory&  Physical quantity & Limit: charge density, $\rho$,\\
 dimension&      at the horizon& and magnetic field, $B$  \\
\hline\hline
$2+1$ dimension & $Re\sigma^{xx}\sim T^{-2/z},~Re\sigma^{xy}\sim T^{-4/z}$ & $\rho$=large and $B$=small \\
\hline
$2+1$ dimension & $Re\sigma^{xx}\sim T^{2/z},~Re\sigma^{xy}\sim$ constant & $\rho$=small   and $B$=large \\
\hline
$2+1$ dimension & $Re\sigma^{xx}\sim$ constant ,~$Re\sigma^{xy}\sim T^{-4/z}$  & $\rho$=small,  $B$=small and \\
&&high temperature\\
\hline
\hline
$3+1$ dimension & $Re\sigma^{xx}\sim T^{-2/z},~ Re\sigma^{yy}\sim T^{-2/z}$& Large charge density   \\
&$Re\sigma^{yz}\sim T^{-4/z}$ &and small magnetic field \\
\hline
$3+1$ dimension & $Re\sigma^{xx}\sim T^{-1/z},~Re\sigma^{yy}\sim T^{3/z}$ & Small charge density   \\
&$Re\sigma^{yz}\sim$ constant &and large magnetic field \\
\hline
$3+1$ dimension & $Re\sigma^{yy}\sim T^{1/z}$  ,~$Re\sigma^{yz}\sim T^{-4/z}$  & $\rho$=small,  $B$=small and \\
&$Re\sigma^{xx}\sim T^{1/z}$ &high temperature\\
\hline
\hline
\end{tabular}
\begin{center}
\noindent 
Table 1: Temperature dependences of the conductivities in $2+1$ and $3+1$ dimensional field theories with dynamical exponent $z$.
\end{center}

In a recent paper \cite{kkp}, the authors have obtained the result of the resistivity of FL and NFL state but there exists few important differences with this paper. (1) We consider strictly unit dynamical exponent. Moreover, we do not need to boost it. (2) The proper treatment is done in the sense that it is the resistivity along the direction of the magnetic field, i.e. $1/\sigma^{xx}$, that interpolates between the FL and NFL state by doing appropriate fine tuning. However,  this component of the resistivity  does not show the required upturn behavior of the resistivity. (3) Here we have used the flow equation technique to calculate the conductivity as suggested for the  Maxwell action in \cite{il}, and done for the DBI action in \cite{ssp1}. In this approach, it is the regularity condition on the conductivity at the horizon  determines the dc conductivity. It means at zero frequency  the conductivity at the boundary is precisely determined by its value at the horizon.  

The paper is organized as follows.
In section 2, we shall calculate the flow equation of the conductivity using the DBI action with non-trivial  embedding field and discuss the ambiguity to set the boundary condition. However, for trivial embeddings there does not exist any ambiguity to fix the boundary condition. Then we find the diffusion constant in arbitrary spacetime  dimensions. We show the existence of a non-trivial diffusion constant even for  zero charge density but with non-zero magnetic field. Then we calculate the flow equation for the conductivity in $2+1$ dimensional field theory by considering the DBI and Chern-Simon action. In section 3, we show the approximate strange metallic behavior for unit dynamical exponent in $3+1$ dimensional field theory but in a different regime of the parameter space, namely,  small charge density and strong magnetic field regime.

\subsection{The approach}

In this section, we shall spell out the approach that we are going to adopt to do the calculations. A method  was advocated in \cite{kob} to compute the dc conductivity in the probe brane approximation. This particular approach is useful only when the  action is of the DBI type. In this setting, the basic bi-fundamental degrees of freedom are moving in a heat bath in the presence of an electric field, $E_j$, along the j'th direction and the motion of the charge carriers generates the current, $J^i$. More importantly, the calculation is done in a limit where the degrees of freedom do not back react on the heat bath \cite{kk}. Then the conductivity is calculated using the Ohm's law, which finally  takes the following form 
\be
J^i=\sigma^{ij}(E,\mu_k)~E_j,
\ee
where the conductivity is a function of the applied electric field and $\mu_k$ denote parameters like temperature, charge density and magnetic field etc., this particular method is used e.g., in \cite{ob},\cite{hpst},\cite{kiritsis},\cite{kachru},\cite{ks},\cite{pal},\cite{kkp},\cite{kim} and \cite{lpp} to do the calculations. Latter, in \cite{il}, the authors gave  a prescription to calculate the  conductivity of the 
Maxwell system in a fixed geometry via the flow equation of the conductivity. According to which, the dc conductivity at the boundary, $r\ra \infty$,   is determined by imposing the regularity condition on the flow equation at the horizon, which is same as putting   the in-falling boundary condition on the gauge field at the horizon. At zero frequency this value of the dc conductivity at the boundary  is determined  by its  value at the horizon. In fact, in the zero frequency limit, it is the same irrespective of the choice of $r$.  In this case the Ohm's law
at a constant slice of $r$ is assumed as 
\be\label{ohms_law_cond_not_E}
J^i=\sigma^{ij}(\mu_k)~E_j,
\ee
where the conductivity is not a function of the applied electric field and the explicit dependence
on $r$ is suppressed. Then the authors of \cite{il} found the flow equation obeyed by the conductivity as one goes from one slice of constant $r$ to another. 
 Following this prescription, in \cite{ssp1}, the authors found the flow equation of the conductivity  for the DBI type of action, among others, using the  Ohm's law  as written in eq(\ref{ohms_law_cond_not_E}) for a constant slice of $r$.
It is shown in \cite{ssp1} that  to leading order in the electric field, the conductivities calculated using  both the approaches as mentioned above give the same answer.

Using the latter approach, we have defined the   magnetic version of the Ohm's law at a constant slice of $r$ as
\be
J^i=\sigma^{ij}_m(\mu_k)~B_j,
\ee
where $\sigma^{ij}_m$ is the magnetic conductivity and $B_j$ is the magnetic field along the j'th direction, then we find the flow equation obeyed by the magnetic conductivity by going from one slice of constant $r$ to another. We mention, in passing, that there exists yet another approach to understand the NFL state, \cite{liu}, \cite{filmv},\cite{fp},\cite{ss}, which is completely different from the previous approaches in a very important way: the basic degrees of  freedom considered to be fermionic in nature. In this paper we do not have anything to offer on this approach except saying that in \cite{filmv}, the authors   have found the strange metallic properties with unit dynamical exponent.

\section{Flow equation with DBI action  in a fixed geometry  }

In this section, we shall write down  the flow equation for the electrical conductivity in a fixed background geometry i.e. in a probe brane approximation. In order to do the calculation, we  consider  the  action to be  of the DBI type. We {\it emphasize}, the flow equation that follows from it is completely different than that found in \cite{il}, simply,  because (a) the actions are different and (b) with DBI action, in the probe brane approximation, the flow equation depends on the charge density whereas for Maxwell type of action it does not.  

Generically, we consider the background geometry as   a  $d+2$ dimensional bulk spacetime, with a trivial dilaton profile, $\phi=0$. We shall consider two different cases to the background geometry. Those are
\bea\label{geometry_fixed}
ds^2_{I,d+2}&=&G_{MN}dx^Mdx^N
=-g_{tt}(r,z)dt^2+g_{xx}(r,z)dx^2+g_{yy}(r,z)dy^2+g_{zz}(r,z)
dz^2+\nn&&\sum^{d-3}_{i=1}g_{ij}(r,z)
dw^idw^j+g_{rr}(r,z)dr^2,\nn
ds^2_{II,d+2}&=&
G_{MN}dx^Mdx^N=-g_{tt}(r)dt^2+g_{xx}(r)dx^2+g_{yy}(r)dy^2+g_{zz}(r)
dz^2+\nn&&\sum^{d-3}_{i=1}g_{ij}(r)
dw^idw^j+g_{rr}(r)dr^2.
\eea
The  DBI action is described by
\be
S=-T_b\int \sqrt{-det([g]+\lambda F)_{ab}}=-T_b\int \sqrt{-det(\p_aX^M\p_b X^N G_{MN}+\lambda F_{ab})},
\ee
where $[g]_{ab}$ is the induced metric on the brane world volume, $T_b$ is the tension of the brane and $\lambda$ is a dimension full object which we shall set  to unity for convenience. The dilaton is assumed to be trivial as we want the scaling symmetry of the (closed string)  background fields. The fields $X^M$
are the map from the brane world volume to the spacetime and are called as the embedding functions.

Let us consider a situation where the probe brane is $d$ dimensional, i.e. it is extended along $x,~y,~w^i$ and $r$.
So, that the $d+1$ dimensional  induced metric  in the static gauge choice becomes
\bea\label{induced_geometry}
ds^2_{I}(ind)&=&-g_{tt}(r,z(r))dt^2+g_{xx}(r,z(r))dx^2+g_{yy}(r,z(r))dy^2+\sum^{d-3}_{i=1}g_{ij}(r,z(r))
dw^idw^j+\nn&&[g_{zz}(r,z(r))z'^2+g_{rr}(r,z(r))]dr^2,\nn
ds^2_{II}(ind)&=&-g_{tt}(r)dt^2+g_{xx}(r)dx^2+g_{yy}(r)dy^2+\sum^{d-3}_{i=1}g_{ij}(r)
dw^idw^j+[g_{zz}(r)z'^2+g_{rr}(r)]dr^2,\nn
\eea
where we have excited the scalar field $z(r)$ and prime, $'$, denotes derivative with respect to $r$. Let us embed this system in an electric and constant magnetic field.
\be
F^{(0)}=F^{(0)}_{rt}(r)dr\w dt+F^{(0)}_{xy}dx\w dy
\ee

Since, we are working in the probe brane approximation, it is consistent to   do an infinitesimal fluctuation to the gauge fields only. 
\bea\label{gauge_field_fixed}
A_a\ra A^{(0)}_a+A^{(1)}_a,\quad\quad a=(t,~x,~y,~i=w^i,~r)
\eea
where the fluctuations are expressed by  $(1)$ in the superscript of the field and we allow the fluctuating fields to depend\footnote{In principle, we are supposed to make the fluctuating fields to depend on all the world volume coordinates, but latter we shall choose the momentum as $k^a=(\omega,k,0,0,...,0)$. In which case, in the Fourier space the only relevant quantity that left over is same as if the fluctuating fields depend on $t,~x,$ and $r$.} on $t,~x,$ and $r$. In  what follows, we have set the gauge choice, $A_r=0$, which means $F^{(0)}_{rt}=A'^{(0)}_t$ and  considered $F^{(0)}_{xy}=B$, to be a constant.

\subsection{Case I: $g_{ab}(r,~z)$}

We shall consider the general case where the induced metric components depend both on the radial direction, $r$, and the  embedding field, $z$, which we denote as case I. From which, it is easy to derive the result for the case II, i.e. by simply taking the metric components to be independent of $z$.

In this case the  solution of the gauge field $A^{(0)}_t(r)$ is
\be
A'^{(0)}_t(r)= \f{\rho~\sqrt{g_{tt}(g_{rr}+g_{zz}z'^2)}}{\sqrt{\rho^2+(\prod_i g_{w^iw^j})(g_{xx}g_{yy}+B^2)}}.
\ee

Analytically, it is very difficult to find the solution to the embedding field $z(r)$, whose equation of motion takes the following form
\bea
&&\p_r\Bigg[\f{\sqrt{g_{tt}}g_{zz}z'{\sqrt{\rho^2+(\prod_i g_{w^iw^j})(g_{xx}g_{yy}+B^2)}}}{\sqrt{g_{rr}+g_{zz}z'^2}} \Bigg]-\nn&&\p_z\Bigg[\sqrt{g_{tt}(g_{rr}+g_{zz}z'^2)}\sqrt{\rho^2+(\prod_i g_{w^iw^j})(g_{xx}g_{yy}+B^2)}\Bigg]=0
\eea

When the metric components are functions of the radial coordinate,  $r$, only, then the solution to the embedding function is
\be\label{cyclic_z}
z'(r)= \f{c ~\sqrt{g_{rr}}}{\sqrt{g_{zz}}\sqrt{g_{tt}g_{zz}[\rho^2+(\prod_i g_{w^iw^j})(g_{xx}g_{yy}+B^2)]-c^2}},
\ee
where $c$ and $\rho$ are the integration constants associated to the  fields $z$ and $A^{(0)}_t$,  and we interpret these constants as the condensate and the charge density, respectively. Since, we are interested to do the calculations, analytically, means we shall consider the metric components to be function of the radial coordinate, $r$, only.

Using the explicit background geometry eq(\ref{induced_geometry}) and
gauge field eq(\ref{gauge_field_fixed}), results in the  fluctuating current\footnote{The general form of the equations of motion and the current associated to the fluctuating fields are written down in \cite{ssp1}.} at a constant slice of $r$ as
\bea\label{current_dbi_b}
J^{(1)x}&=&-T_b\sqrt{\prod^{d-3}_{i=1} g_{w^iw^j}}\Bigg(\f{g_{tt}g_{yy}\p_r A^{(1)}_x-i\omega B \p_rA^{(0)}_t A^{(1)}_y}{\sqrt{g_{tt}(g_{rr}+g_{zz}z'^2)-(\p_r A^{(0)}_t)^2}\sqrt{g_{xx}g_{yy}+B^2}}\Bigg)\nn
J^{(1)y}&=&-T_b\sqrt{\prod^{d-3}_{i=1} g_{w^iw^j}}\Bigg(\f{g_{tt}g_{xx}\p_r A^{(1)}_y+iB\p_rA^{(0)}_t E^{(1)}}{\sqrt{g_{tt}(g_{rr}+g_{zz}z'^2)-(\p_r A^{(0)}_t)^2}\sqrt{g_{xx}g_{yy}+B^2}}\Bigg)\nn
J^{(1)i}&=&-T_b\sqrt{\prod^{d-3}_{i=1} g_{w^iw^j}}
\Bigg(\f{g_{tt}\sqrt{g_{xx}
g_{yy}+B^2}~g^{ij}\p_r A^{(1)}_j}{\sqrt{ g_{tt}(g_{rr}+g_{zz}z'^2)-(\p_r A^{(0)}_t)^2}}\Bigg),
\eea
where we have done the Fourier transformation of the (fluctuating)  field  with the convention of $e^{i(kx-wt)}$ and $E^{(1)}\equiv\omega A^{(1)}_x+kA^{(1)}_t $. The inverse of  $g_{ij}$ is denoted as $g^{ij}$, i.e. $g^{ik}g_{kj}=\delta^i_j$. From now on, we shall set the tension of the brane to unity, $T_b=1$, for simplicity. The explicit form of the equations of motion of the gauge fields
\bea
&&\p_r\Bigg[\sqrt{\prod_i g_{w^iw^j}}\Bigg(\f{g_{tt}g_{yy}\p_r A^{(1)}_x-i\omega B \p_rA^{(0)}_t A^{(1)}_y}{\sqrt{g_{tt}(g_{rr}+g_{zz}z'^2)-(\p_r A^{(0)}_t)^2}\sqrt{g_{xx}g_{yy}+B^2}}\Bigg) \Bigg]+\nn
&&\omega ~
\sqrt{\prod_i g_{w^iw^j}}\Bigg(\f{(g_{rr}+g_{zz}z'^2)g_{yy} E^{(1)}+i B \p_rA^{(0)}_t \p_rA^{(1)}_y}{\sqrt{g_{tt}(g_{rr}+g_{zz}z'^2)-(\p_r A^{(0)}_t)^2}\sqrt{g_{xx}g_{yy}+B^2}}\Bigg)=0,
\eea

\bea
&&\p_r\Bigg[\f{\sqrt{\prod_i g_{w^iw^j}}\sqrt{g_{xx}g_{yy}+B^2}g_{tt}(g_{rr}+g_{zz}z'^2)\p_r A^{(1)}_t}{[g_{tt}(g_{rr}+g_{zz}z'^2)-(\p_r A^{(0)}_t)^2]^{3/2}}+\f{\sqrt{\prod_i g_{w^iw^j}}ik B \p_rA^{(0)}_t A^{(1)}_y}{\sqrt{g_{tt}(g_{rr}+g_{zz}z'^2)-(\p_r A^{(0)}_t)^2}
\sqrt{g_{xx}g_{yy}+B^2}} \Bigg]\nn
&&-k~
\sqrt{\prod_i g_{w^iw^j}}\Bigg(\f{(g_{rr}+g_{zz}z'^2)g_{yy} E^{(1)}+i B \p_rA^{(0)}_t \p_rA^{(1)}_y}{\sqrt{g_{tt}(g_{rr}+g_{zz}z'^2)-(\p_r A^{(0)}_t)^2}\sqrt{g_{xx}g_{yy}+B^2}}\Bigg)=0,
\eea

\bea
&&\p_r\Bigg[\sqrt{\prod_i g_{w^iw^j}}\Bigg(\f{g_{tt}g_{xx}\p_r A^{(1)}_y+iB \p_r A^{(0)}_t E^{(1)}}{\sqrt{g_{tt}(g_{rr}+g_{zz}z'^2)-(\p_r A^{(0)}_t)^2}\sqrt{g_{xx}g_{yy}+B^2}}\Bigg) \Bigg]+\f{\sqrt{\prod_i g_{w^iw^j}}}{\sqrt{g_{xx}g_{yy}+B^2}}\times \nn
&&\Bigg[\f{\omega^2g_{xx}(g_{rr}+g_{zz}z'^2)
A^{(1)}_y-iB \p_r A^{(0)}_t \p_r E^{(1)} }{\sqrt{g_{tt}(g_{rr}+g_{zz}z'^2)-(\p_r A^{(0)}_t)^2}}-
\f{g_{xx}g_{yy} k^2\sqrt{g_{tt}(g_{rr}+g_{zz}z'^2)-(\p_r A^{(0)}_t)^2}A^{(1)}_y}{g_{xx}g_{yy}+B^2}\Bigg]=0,\nn&&
\p_r\Bigg[\f{\sqrt{\prod_i g_{w^iw^j}}
\sqrt{g_{xx}g_{yy}+B^2}g_{tt}
g^{ij}\p_r A^{(1)}_j}{\sqrt{g_{tt}(g_{rr}+g_{zz}z'^2)-(\p_r A^{(0)}_t)^2}} \Bigg]+\f{\sqrt{\prod_i g_{w^iw^j}}
\sqrt{g_{xx}g_{yy}+B^2}}{\sqrt{g_{tt}(g_{rr}+g_{zz}z'^2)-(\p_r A^{(0)}_t)^2}}g_{rr}g^{ij}
\omega^2 A^{(1)}_j\nn&&-\f{\sqrt{\prod_i g_{w^iw^j}}\sqrt{g_{tt}(g_{rr}+g_{zz}z'^2)-(\p_r A^{(0)}_t)^2}}{{\sqrt{g_{xx}g_{yy}+B^2}}}g_{yy}
g^{ij}k^2A^{(1)}_j=0
\eea

and the constraint equation
\be
\omega (g_{rr}+g_{zz}z'^2)(g_{xx}g_{yy}+B^2)\p_rA^{(1)}_t+kg_{yy}[g_{tt}(g_{rr}+g_{zz}z'^2)-(\p_r A^{(0)}_t)^2]\p_r A^{(1)}_x=0.
\ee

It is easy to notice that the gauge field component $A^{(1)}_j$ decouples from the other components of the guage field. So, we can deal with this component of the gauge field independent of others.

\subsection{Zero magnetic field}

Here we shall  find the flow equation of
the conductivity in the zero magnetic field case. It follows trivially from the equation of motion that in the zero magnetic field  limit, the $A^{(1)}_y$ component of the gauge field decouples from the rest. 
In which case, the equation of motion to $E^{(1)}$ becomes
\be
\p^2_r E^{(1)}+\f{\p_r E^{(1)} \p_r {\tilde X}}{{\tilde X}}+\Bigg[\f{\omega^2 g_{xx}(g_{rr}+g_{zz}z'^2)-k^2[g_{tt}(g_{rr}+g_{zz}z'^2)-(\p_r A^{(0)}_t)^2]}{g_{tt}g_{xx}}\Bigg]E^{(1)}=0,
\ee
where the function 
\be
{\tilde X}=\f{\sqrt{(\prod_i g_{w^iw^j})g_{xx}g_{yy}}~g_{tt}(g_{rr}+g_{zz}z'^2)}{\sqrt{g_{tt}(g_{rr}+g_{zz}z'^2)-(\p_r A^{(0)}_t)^2}[\omega^2 g_{xx}(g_{rr}+g_{zz}z'^2)-k^2(g_{tt}(g_{rr}+g_{zz}z'^2)-(\p_r A^{(0)}_t)^2)]}
\ee

Differentiating the Ohm's law, $
J^{(1)x}=\sigma^{xx}F^{(1)}_{xt}=i\sigma^{xx} E^{(1)}$, with respect to $r$ and using the equations of motion of the gauge field, $E^{(1)}$,  follows the flow equation for the conductivity
\bea
\p_r \sigma^{xx}&=&i\omega \Bigg[\f{\sqrt{g_{tt}(g_{rr}+g_{zz}z'^2)-(\p_r A^{(0)}_t)^2}~(\sigma^{xx})^2}{g_{tt}\sqrt{(\prod g_{w^iw^j})g_{xx}g_{yy}} (g_{rr}+g_{zz}z'^2)}\times\nn&&\Bigg(g_{xx}(g_{rr}+g_{zz}z'^2)-\f{k^2}{\omega^2}[g_{tt}(g_{rr}+g_{zz}z'^2)-(\p_r A^{(0)}_t)^2]\Bigg)\nn && -\f{\sqrt{(\prod_i g_{w^iw^j})g_{xx}g_{yy}}~ g_{tt}(g_{rr}+g_{zz}z'^2)}{\sqrt{g_{tt}(g_{rr}+g_{zz}z'^2)-(\p_r A^{(0)}_t)^2}~g_{tt}g_{xx}}\Bigg]
\eea

The following combination of the solution of the  gauge field, $A^{(0)}_t(r)$, and  the embedding field, $z'(r)$, gives
\be
g_{tt}(g_{rr}+g_{zz}z'^2)-(\p_r A^{(0)}_t)^2=\f{g_{tt}(g_{rr}+g_{zz}z'^2)(\prod_i g_{w^iw^j})g_{xx}g_{yy}}{\rho^2+(\prod_i g_{w^iw^j})g_{xx}g_{yy}},
\ee
where $\rho$ is the constant of motion and can be interpreted as the charge density. Substituting this into the flow equation of the  conductivity results in
\bea
\p_r \sigma^{xx}&=&\f{i\omega \sqrt{g_{rr}+g_{zz}z'^2}}{g_{xx}\sqrt{g_{tt}[\rho^2+(\prod_i g_{w^iw^j})g_{xx}g_{yy}]}}\Bigg[(\sigma^{xx})^2 g_{xx}\nn&&\Bigg(g_{xx}-\bigg(\f{k^2}{\omega^2}\bigg)\bigg(\f{g_{tt}(\prod_i g_{w^iw^j})g_{xx}g_{yy}}{\rho^2+(\prod_i g_{w^iw^j})g_{xx}g_{yy}}\bigg)\Bigg)-\Bigg(\rho^2+(\prod_i g_{w^iw^j})g_{xx}g_{yy}\Bigg)\Bigg].
\eea

Let us consider a situation for which  the  combination $g_{rr}+g_{zz}z'^2$  does not vanishes at the horizon, $ (g_{rr}+g_{zz}z'^2)_{r_h}\neq 0$. Then the
regularity condition at the horizon, where $g_{tt}(r_h)=0$,  requires that we need to impose the following boundary condition
\be\label{regularity_condition_B_0}
(\sigma^{xx})^2_{r_h}=\Bigg[\f{\rho^2+(\prod_i g_{w^iw^j})g_{xx}g_{yy}}{g^2_{xx}} \Bigg]_{r_h}.
\ee

 In the high density limit, the dc  conductivity at the horizon  approximately becomes,  $\sigma^{xx}(r_h)\sim \f{\rho}{g_{xx}(r_h)}$. This form of the dc conductivity, essentially, suggests that it's form remain unchanged even for non-trivial embeddings.

\subsubsection{Cyclic $z(r)$}
If we consider a situation, i.e. case II, for which the metric components are not functions of $z$, then it means, the momentum associated to the field $z(r)$ is constant. The solution to $z(r)$ can be found by integrating eq(\ref{cyclic_z}). In this case
\be
g_{rr}+g_{zz}z'^2=\f{g_{rr}g_{tt}g_{zz}[\rho^2+(\prod_i g_{w^iw^j})g_{xx}g_{yy}]}{g_{tt}g_{zz}[\rho^2+(\prod_i g_{w^iw^j})g_{xx}g_{yy}]-c^2}.
\ee

In which case, the flow equation of the conductivity becomes
\bea
\p_r \sigma^{xx}&=&\f{i\omega \sqrt{g_{rr}g_{zz}}}{g_{xx}\sqrt{g_{tt}g_{zz}[\rho^2+(\prod_i g_{w^iw^j})g_{xx}g_{yy}]-c^2}}\Bigg[(\sigma^{xx})^2 g_{xx}\nn&&\Bigg(g_{xx}-\bigg(\f{k^2}{\omega^2}\bigg)\bigg(\f{g_{tt}(\prod_i g_{w^iw^j})g_{xx}g_{yy}}{\rho^2+(\prod_i g_{w^iw^j})g_{xx}g_{yy}}\bigg)\Bigg)-\Bigg(\rho^2+(\prod_i g_{w^iw^j})g_{xx}g_{yy}\Bigg)\Bigg].
\eea
In this case, the choice of the boundary condition is a bit subtle. If we choose a coordinate system for which $g_{rr}$ diverges at the horizon, $r_h$, then the regularity at the horizon yields the same condition on the conductivity as in eq(\ref{regularity_condition_B_0}). 
Note that for non-zero condensate, $c\neq 0$, there exists a value of the radial coordinate,  $r_{\star}$, for which
\be
[\rho^2+(\prod_i g_{w^iw^j})g_{xx}g_{yy}]_{r_{\star}}=
c^2/(g_{tt}g_{zz})_{r_{\star}},
\ee
in which case, it follows that the solution to $A^{(0)}_t(r)$,  the embedding field, $z'(r)$, and the
induced metric are singular i.e. not well behaved. If  the radial value $r_{\star}$ is not a physical scale in the sense that it stays inside the horizon then all the fields are well behaved.  However, if $r_{\star}>r_h$, then 
such a condensate is not allowed. So, for the DBI action and with zero condensate there is not any ambiguity to set the boundary condition. However, that is not the case for non-trivial embeddings,  i.e. with non-zero condensate.

\subsection{Electric charge diffusion}

 The diffusion constant can be derived from the flow equation of the conductivity in the hydrodynamic limit \cite{il}. This is, essentially,
 studied from the gravity by going over to the hydrodynamic regime, where $\omega\sim k^2\ra 0,~ \omega\ll T$ and $k\ll T$.  In which case, we can rewrite the flow equation of the conductivity as
\bea
\f{\p_r \sigma^{xx}}{T}&=&\f{i{\hat\omega} \sqrt{(g_{rr}+g_{zz}z'^2)}}{g_{xx}\sqrt{g_{tt}[\rho^2+(\prod_i g_{w^iw^j})g_{xx}g_{yy}]}}\Bigg[(\sigma^{xx})^2 g_{xx}\nn&&\Bigg(g_{xx}-\bigg(\f{{\hat k}^2}{{\hat\omega}^2}\bigg)\bigg(\f{g_{tt}(\prod_i g_{w^iw^j})g_{xx}g_{yy}}{\rho^2+(\prod_i g_{w^iw^j})g_{xx}g_{yy}}\bigg)\Bigg)-\Bigg(\rho^2+(\prod_i g_{w^iw^j})g_{xx}g_{yy}\Bigg)\Bigg],
\eea
where ${\hat\omega}=\omega/T$ and  ${\hat k}=k/T$.  Considering $\sigma^{xx}\sim {\cal O}(1) $, and for small density we can rewrite it as
\be
\f{\p_r \sigma^{xx}}{T}\simeq -i\bigg(\f{{\hat k}^2}{{\hat\omega}}\bigg)\f{ \sqrt{g_{tt}(g_{rr}+g_{zz}z'^2)}(\sigma^{xx})^2(\prod_i g_{w^iw^j})g_{xx}g_{yy}}{[\rho^2+(\prod_i g_{w^iw^j})g_{xx}g_{yy}]^{3/2}}
\ee

Now, integrating it from the horizon, $r_h$, to some generic point,  $r$
\be
\f{1}{\sigma^{xx}(r)}=\f{1}{\sigma^{xx}(r_h)}+i\f{k^2}{\omega}\int^r_{r_h} dr' \f{ \sqrt{g_{tt}(g_{rr}+g_{zz}z'^2)}(\prod_i g_{w^iw^j})g_{xx}g_{yy}}{[\rho^2+(\prod_i g_{w^iw^j})g_{xx}g_{yy}]^{3/2}}.
\ee

The retarded  Green function at the boundary, defined as,  $G^{xx}_R(\omega,~k)=-i\omega \sigma^{xx}(r\rightarrow \infty)$ gives
\be\label{diff_def}
G^{xx}_R(\omega,~k)=\f{\omega^2 \sigma^{xx}(r_h)}{i\omega-Dk^2},
\ee
where the diffusion constant at the boundary  is
\be
D= \sigma^{xx}(r_h) \int^{\infty}_{r_h}dr'
\f{ \sqrt{g_{tt}(g_{rr}+g_{zz}z'^2)}~(\prod_i g_{w^iw^j})g_{xx}g_{yy}}{[\rho^2+(\prod_i g_{w^iw^j})g_{xx}g_{yy}]^{3/2}}.
\ee

Now using the Einstein relation, $ \Xi=\sigma^{xx}(r_h)/D$, with $\Xi$ is the charge susceptibility
\be
\Xi=\Bigg( \int^{\infty}_{r_h}dr'
\f{ \sqrt{g_{tt}(g_{rr}+g_{zz}z'^2)}~(\prod_i g_{w^iw^j})g_{xx}g_{yy}}{[\rho^2+(\prod_i g_{w^iw^j})g_{xx}g_{yy}]^{3/2}} \Bigg)^{-1}.
\ee

Eq(\ref{diff_def}) suggests that the Green function,  $G^{xx}_R(\omega,~k)$, in the hydrodynamic limit  has a pole
\be\label{dispersion}
\omega=-iD k^2.
\ee

In the zero density limit and for trivial embedding, as expected, we do reproduce the diffusion constant for the Maxwell system. The dispersion relation in the hydrodynamic limit, eq(\ref{dispersion}), suggests the appearance of  a gapless spectrum at zero momentum.

\subsection{Diffusion for zero charge density and with non-trivial embedding}

The  equation of motion to $E^{(1)}$ in the absence of the  charge density, $\rho=0$, with non-trivial embedding and non-zero magnetic field, $B\neq 0$, is
\be
\p^2_r E^{(1)}+\f{\p_r E^{(1)} \p_r {\tilde X}}{{\tilde X}}+(g_{rr}+g_{zz}z'^2)\Bigg[\f{\omega^2 (g_{xx}g_{yy}+B^2)-k^2 g_{yy}g_{tt}]}{g_{tt}(g_{xx}g_{yy}+B^2)}\Bigg]E^{(1)}=0,
\ee
where
\be
{\tilde X}=\f{\sqrt{(\prod_i g_{w^iw^j})g_{tt}}~g_{yy}\sqrt{g_{xx}g_{yy}+B^2}}{\sqrt{(g_{rr}+g_{zz}z'^2)}[\omega^2 (g_{xx}g_{yy}+B^2)-k^2g_{tt}g_{yy}]}.
\ee

The current, ${J^{(1)}}^x$, at a constant slice of $r$ can be re-written as
\be
{J^{(1)}}^x=-\omega \f{\sqrt{(\prod_i g_{w^iw^j})g_{tt}}~g_{yy}\sqrt{(g_{xx}g_{yy}+B^2)}}{\sqrt{(g_{rr}+g_{zz}z'^2)}[\omega^2 (g_{xx}g_{yy}+B^2)-k^2g_{tt}g_{yy}]} \p_r E^{(1)}=-\omega {\tilde X}\p_r E^{(1)}
\ee

Now, using the Ohm's law, ${J^{(1)}}^x=i\sigma^{xx}E^{(1)}$, we get the following flow equation of the conductivity
\bea
\p_r \sigma^{xx}&=&\f{i\omega\sqrt{g_{rr}+g_{zz}z'^2}}{\sqrt{g_{tt}(\prod_i g_{w^iw^j})}\sqrt{g^2_{yy}(g_{xx}g_{yy}+B^2)}}\Bigg[(\sigma^{xx})^2\bigg( g_{xx}g_{yy}+B^2-\f{k^2}{\omega^2}g_{yy}g_{tt}\bigg)\nn&&- (\prod_i g_{w^iw^j})g^2_{yy}\Bigg].
\eea

Assuming that the quantity, $g_{rr}+g_{zz}z'^2$, does not vanish at the horizon, $r_h$, gives us the necessary regularity condition which needed to be imposed on the conductivity at the horizon. It reads as
\be
\sigma^{xx}(r_h)=\Bigg(\f{\sqrt{(\prod_ig_{w^iw^j})}~g_{yy}}
{\sqrt{g_{xx}g_{yy}+B^2}}\Bigg)_{r_h}.
\ee

In the hydrodynamic limit, $\omega\sim k^2<<T$,  we can find the Green function at the boundary, $G^{xx}_R(\omega,~k)=-i\omega \sigma^{xx}(r\rightarrow \infty)=\f{\omega^2 \sigma^{xx}(r_h)}{i\omega-Dk^2}$. The quantity $D$ is called as the diffusion constant in the field theory and it takes the following form
\be\label{diff_non_zero_c_b}
D=\sigma^{xx}(r_h)\int^{\infty}_{r_h}dr~
\f{\sqrt{g_{tt}(g_{rr}+g_{zz}z'^2)}}{\sqrt{(\prod_i g_{w^iw^j})(g_{xx}g_{yy}+B^2)}}
\ee

The integral in the diffusion constant can be calculated  explicitly  for trivial embedding,  $z'(r)=0$, i.e. in the zero condensate limit, $c=0$. Evaluating it explicitly for the  asymptotically  $d+1$ dimensional induced AdS  spacetime on the brane world volume, we find
\be
D=\f{L}{(d-2)} \bigg(\f{TL} {\alpha}\bigg)^{-1}\f{1}{\sqrt{1+B^2\bigg(\f{\alpha}{TL}\bigg)^{4}}}{}_2F_1\Bigg[\f{1}{2},\f{d-2}{4},\f{d+2}{4},-B^2\bigg(\f{\alpha}{TL}\bigg)^{4}\Bigg],
\ee
where $\alpha=d/4\pi$ and ${}_2F_1[a,b,c,x]$ is the hypergeometric function.
In passing, let us point out that in the zero charge density, vanishing magnetic field and trivial embedding function limit, the analysis for the DBI action reduces to that of the Maxwell system  in \cite{il}.

\subsection{Flow equation with DBI and Chern-Simon term}

In this subsection, we shall find the flow equation of the conductivity in $2+1$ dimensional field theory with both DBI and CS term in the action. In which case, the bulk action  is
\be
S=-T_3\int \sqrt{-det([g]+F)_{ab}}+\mu_3\int\bigg[
[C_4]+
[C_2] \w F_2+\f{[C_0]}{2} F_2\w F_2\bigg],
\ee
where $C_n$ are the n-form RR potentials and [~] stands for the pull back of the closed string fields onto the world volume of the brane.
Given the closed string fields, the $F_2$ is determined by solving the equation of motion that follows from the action. We assume, for simplicity, $C_4=0=C_2$ and $C_0=\theta$,  which is assumed to be constant. The induced metric with trivial embedding function is assumed as
\be
ds^2_4=-g_{tt}(r)dt^2+g_{xx}(r)[dx^2+dy^2]+g_{rr}(r)dr^2.
\ee
From now on, we shall set $T_3=\mu_3$. The 2-form field strength, $F^{(0)}_2=A'_t(r) dr\w dt+ Bdx\w dy $,  that solves the equation of motion takes the following form
\be
A'^2_t=\f{(\rho-4\theta B)^2 g_{tt}g_{rr}}{(\rho-4\theta B)^2+g^2_{xx}+B^2}.
\ee

Doing the infinitesimal fluctuation to the gauge potential, $A\ra A(r)+A^{(1)}(r,t,x)$, the action to quadratic order in the fluctuation
\bea
S^{(2)}&=&-T_3\int \sqrt{-det M^{(0)}_+}\bigg[-
\f{1}{4}{\bigg(M^{(0)}_+\bigg)^{-1}}^{ab}F^{(1)}_{bc}{\bigg(M^{(0)}_+\bigg)^{-1}}^{cd}F^{(1)}_{da}+
\f{1}{8}\nn&&{\bigg(M^{(0)}_+\bigg)^{-1}}^{ab}F^{(1)}_{ab}{\bigg(M^{(0)}_+
\bigg)^{-1}}^{cd}
F^{(1)}_{cd}\bigg]+\f{\theta}{2}\int \epsilon^{abcd} F^{(1)}_{ab}F^{(1)}_{cd},
\eea
where $M^{(0)}_+=g+F^{(0)}$.
The equation of motion and the current at a constant slice of $r$ that results from it in the Fourier space\footnote{The Fourier transformation is done with respect to $e^{i(kx-\omega t)}$.} with the choice of momentum $k^{\mu}=(\omega,k,0)$ are
\bea
&&\p_r\Bigg[ \f{g_{tt}g_{xx}\p_r A^{(1)}_{\pm}\mp\omega B A'_t A^{(1)}_{\pm}}{\sqrt{g_{tt}g_{rr}-A'^2_t}\sqrt{g^2_{xx}+B^2}}
\Bigg]+\f{[\omega^2g_{rr}
g_{xx}A^{(1)}_{\pm}\pm \omega B A'_t\p_r A^{(1)}_{\pm}]}{\sqrt{g_{tt}g_{rr}-A'^2_t}\sqrt{g^2_{xx}+B^2}}=0,\nn&&J^{(1)}_{\pm}=-T_3\Bigg[\f{g_{tt}g_{xx}\p_r A^{(1)}_{\pm}\mp\omega B A'_t A^{(1)}_{\pm}}{\sqrt{g_{tt}g_{rr}-A'^2_t}\sqrt{g^2_{xx}+B^2}}\mp 4\theta\omega A^{(1)}_{\pm}\Bigg],
\eea
where we have set momentum to $k=0$ and $A^{(1)}_t=0$. Let us define   $A^{(1)}_{\pm}\equiv A^{(1)}_x\pm iA^{(1)}_y$ and $J^{(1)}_{\pm}\equiv J^{(1)}_x\pm iJ^{(1)}_y$, in which case, the Ohm's law  becomes
\be
J^{(1)}_{\pm}=\pm\omega \sigma_{\pm }A^{(1)}_{\pm}, \quad \quad \sigma_{\pm }\equiv \sigma^{xy}\pm i\sigma^{xx}.
\ee

 Using the equation of motion of $A^{(1)}_{\pm}$, there follows the flow equation for the conductivity
\bea
\p_r \sigma_{\pm}&=&\mp \f{2\omega B A'_t}{g_{tt}g_{xx}}(\sigma_{\pm}-4\theta T_3)\pm \f{\omega\sqrt{g_{tt}g_{rr}-A'^2_t}\sqrt{g^2_{xx}+B^2}}{T_3g_{tt}g_{xx}}(\sigma_{\pm}-4\theta T_3)^2\pm\nn&& \f{T_3\omega}{\sqrt{g_{tt}g_{rr}-A'^2_t}\sqrt{g^2_{xx}+B^2}}
\Bigg(g_{tt}g_{rr}+\f{B^2A'^2_t}{g_{tt}g_{xx}}\Bigg).
\eea
The explicit form of the flow equation of $\sigma^{xy}$ and $\sigma^{xx}$ using the solution
of $A'_t$
\bea
\p_r\sigma^{xx}&=&-
\f{i\omega
\sqrt{g_{rr}}}{T_3 g_{xx}\sqrt{g_{tt}}
\sqrt{(\rho-4\theta B)^2+B^2+g^2_{xx}}}\times\Bigg[
T^2_3[(\rho-4\theta B)^2+g^2_{xx}]-\nn&&2T_3(\rho-4\theta B) B(\sigma^{xy}-4\theta T_3)+(g^2_{xx}+B^2)\bigg((\sigma^{xy}-4\theta T_3)^2-(\sigma^{xx})^2\bigg)\Bigg],\nn
\p_r\sigma^{xy}&=&-
\f{2i\omega\sigma^{xx}
\sqrt{g_{rr}}[T_3(\rho-4\theta B) B-(g^2_{xx}+B^2)(\sigma^{xy}-4\theta T_3)]}{T_3 g_{xx}\sqrt{g_{tt}}
\sqrt{(\rho-4\theta B)^2+B^2+g^2_{xx}}}.
\eea
The regularity condition at the horizon, $r_h$, requires us to set
\be\label{bc_sigma_db_cs_2+1_d}
Re\sigma^{xx}(r_h)= T_3
\Bigg[\f{g_{xx}\sqrt{(\rho-4\theta B)^2+B^2+g^2_{xx}}}{B^2+g^2_{xx}}\Bigg]_{r_h},~ Re\sigma^{xy}(r_h)=\f{T_3(\rho-4\theta B) B}{g^2_{xx}(r_h)+B^2}+4\theta T_3,
\ee
while the imaginary part of the conductivity vanishes. This form of the dc conductivity precisely matches with that written in \cite{ssp}, up to a redefinition of the axion and with \cite{akvk} up to a redefinition of the charge density. As an aside, let us assume the spatial part of the metric component as, $g_{xx}(r_h)=r^2_h\sim T^{2/z}$, where $T$ is the Hawking temperature and $z$ the dynamical exponent. In which case, the quantity
\be
\f{\p R_{xy}(r_h)}{\p B}=\f{(\rho-4\theta B) (B^2- T^{4/z})}{T_3[(\rho-4\theta B) B+4\theta(T^{4/z}+B^2)^2]},
\ee
where the resistivity is defined as $R_{xy}=1/\sigma^{xy}$ and $\theta$ is assumed to be independent of $B$.  Let us define a limit where the charge density is large, small magnetic field, $\theta$ large such that  $\rho B=$ small  and $\theta B=$small, in which case, we can approximate
\be
\f{\p R_{xy}(r_h)}{\p B}\simeq -\f{\rho}{4\theta T_3} T^{-4/z}\quad {\rm for}\quad B^2 < T^{4/z}.
\ee
However, for $B^2> T^{4/z}$, we get
\be
\f{\p R_{xy}(r_h)}{\p B}\simeq \f{\rho B^2}{4\theta T_3} T^{-8/z}.
\ee

Similarly, in the large charge density limit and small but finite magnetic field limit, the quantity
\be
\f{\p R_{xx}(r_h)}{\p B}=\f{B[2(\rho-4\theta B)^2+B^2+T^{4/z}]}{T^{2/z}[(\rho-4\theta B)^2+B^2+T^{4/z}]^{3/2}}\ra \f{2B}{\rho}T^{-2/z}.
\ee

\section{Strange metallic behavior with $z=1$}

In this section, we shall show the appearance of the strange metallic behavior  for asymptotically AdS spacetime using the prescription of \cite{kob}. Here the strange metallic behavior  mean  the following temperature dependence of the longitudinal electrical conductivity: $\sigma\sim T^{-1}$.   This  feature  was previously shown in a well designed setting with AdS/CFT arguments but with dynamical exponent, $z=2$ \cite{hpst}. The spacetime that shows this particular value of the dynamical exponent are called as Lifshitz spacetime in \cite{klm}. The requirement to see such a behavior of the conductivity is to consider a very high charge density system in any dimensional field theory \cite{ssp1}. In what follows, we shall show the same feature of the conductivity in  $3+1$ dimensional field theory using the AdS/CFT arguments  with unit dynamical exponent, $z=1$,  at small charge density and at strong magnetic field limit.

The action that we shall consider is of the DBI type. We assume the metric is diagonal and the 2-form field strength has the following structure
\bea
ds^2&=&-g_{tt}(r)dt^2+g_{xx}(r)dx^2+g_{yy}(r)dy^2+g_{zz}(r)dz^2+g_{rr}(r)dr^2\nn
F_2&=&-E dt\w dx+B dy\w dz+A'_t(r) dr\w dt-H'(r) dr\w dx.
\eea

With this form of the metric and the U(1) gauge field, the DBI action becomes
\bea
S&=&-T_b\int\sqrt{-det([g]+F)_{ab}}\nn
&=&-T_b\int\sqrt{g_{yy}g_{zz}+B^2}\sqrt{g_{tt}g_{rr}g_{xx}-g_{rr}E^2+g_{tt}H'^2(r)-g_{xx}A'^2_t},
\eea

where $[~]$ corresponds to the pull-back of the spacetime metric onto the world volume of the brane and $T_b$ stands for the tension of the brane. Since, the action does not depend on the fields like $H(r)$ and $A_t(r)$, suggests that the corresponding momenta are constants. Let us denote those constants as $C_H$ and $C_{A_t}$, respectively.
\bea
C_H&=&-\f{T_bg_{tt}H'\sqrt{g_{yy}g_{zz}+B^2}}{\sqrt{g_{tt}g_{rr}g_{xx}-g_{rr}E^2+g_{tt}H'^2(r)-g_{xx}A'^2_t}},\nn
C_{A_t}&=&\f{T_bg_{xx}A'_t\sqrt{g_{yy}g_{zz}+B^2}}{\sqrt{g_{tt}g_{rr}g_{xx}-g_{rr}E^2+g_{tt}H'^2(r)-g_{xx}A'^2_t}}.
\eea

From these two equations, there follows the relation and the solution
\bea
H'^2(r)&=&\f{C^2_H}{C^2_{A_t}} \f{g^2_{xx}(r)}{g^2_{tt}(r)}A'^2_t(r)\nn
A'^2_t(r)&=&\f{g_{rr}g_{tt}(g_{tt}g_{xx}-E^2)~C^2_{A_t}}{T^2_b g_{tt}g^2_{xx}(g_{yy}g_{zz}+B^2)-g_{xx}(C^2_H g_{xx}-C^2_{A_t}g_{tt})}.
\eea

Now, we can calculate the Legendre transformed action
\bea
S_L&=&S-\int \f{\delta S}{\delta H'}H'- \int\f{\delta S}{\delta A'_t}A'_t
=-T_b\int\f{\sqrt{g_{yy}g_{zz}+B^2}
[g_{rr}(g_{tt}g_{xx}-E^2)]}{\sqrt{g_{tt}g_{rr}g_{xx}-
g_{rr}E^2+g_{tt}H'^2(r)-g_{xx}A'^2_t}}\nn
&=&T_b \int \sqrt{\f{g_{rr}}{g_{tt}g_{xx}}}\sqrt{(g_{tt}g_{xx}-E^2)[ g_{tt}g_{xx}(g_{yy}g_{zz}+B^2)+c^2_{A_t}g_{tt}-c^2_H g_{xx}]},
\eea
where in the last line, we have used the solution to $H(r)$, and $A_t(r)$, Also, we have  redefined the constants as $C_H=T_b c_H$ and $C_{A_t}=T_bc_{A_t}$.   Now, we can use the arguments of the reality of the Legendre transformed action as in \cite{kob}  to find the constraints at $r_{\star}$ as
\bea
&&g_{tt}(r_{\star})g_{xx}(r_{\star})=E^2\nn
c^2_H&=&\bigg[\f{g_{tt}g_{xx}(g_{yy}g_{zz}+B^2)+c^2_{A_t}g_{tt}}{g^2_{xx}}\bigg]_{r_{\star}}=E\bigg[\f{c^2_{A_t}+g_{xx}(g_{yy}g_{zz}+B^2)}{g^2_{xx}}\bigg]_{r_{\star}},
\eea
where in the second equality of the second line, we have used the first equation. Note that the constant $c_H$ is nothing but the current  $J_x$. Now, using the Ohm's law $
J_x=\sigma E$, gives the conductivity as
\be\label{special_long_cond}
\sigma=\bigg[\f{\sqrt{c^2_{A_t}+g_{xx}(g_{yy}g_{zz}+B^2)}}{g_{xx}}\bigg]_{r_{\star}}.
\ee

This is our master equation for  this section. Let us assume that the charge density, $c_{A_t}$, is very large in comparison to the magnetic field, which means, the first term in the square-root of the conductivity dominates over the second term. In which case, the conductivity reduces to
\be
\sigma\sim c_{A_t}/g_{xx}(r_{\star}).
\ee

Let us consider a spacetime that asymptotes to Lifshitz spacetime means $g_{xx}(r_{\star})= r^2_{\star}\sim T^{2/z}$, where $T$ is the Hawking temperature. In this case, by using the previous formula gives us the necessary non-Fermi liquid like behavior of the conductivity for $z=2$. Let us look at a different corner of the parameter space, where the density is small but the magnetic field is very large. Essentially, it is the last term in the square-root  of eq(\ref{special_long_cond}) dominates. In which case, the conductivity reduces to
\be
\sigma\sim \f{B}{\sqrt{g_{xx}(r_{\star})}}\sim \f{B}{T^{1/z}},
\ee
which for $z=1$ gives us the sought after  behavior of the conductivity. This happens for the asymptotically AdS spacetime but with small charge density and high magnetic field limit.

To visualize things in a nice way, let us rewrite eq(\ref{special_long_cond}) in terms of the parameters like temperature, magnetic field and charge density. To leading order in the electric field, it becomes
\be\label{master_formula_cond}
\sigma\simeq \sqrt{c^2_{A_t}T^{-4/z}+T^{2/z}+B^2 T^{-2/z}},
\ee
where we have assumed the presence of a rotational symmetry in the background geometry. From this expression, it is easy to see that at low temperature and small magnetic field limit the conductivity reduces to
\be\label{fl_cond}
\sigma\sim c_{A_t}~T^{-2/z},\quad {\rm For}\quad  (B~,T)\ra 0.
\ee
Whereas at low temperature and  very small charge density limit, the conductivity reduces to
\be\label{nfl_cond}
\sigma\sim B~T^{-1/z},\quad {\rm For}\quad  (c_{A_t}~,T)\ra 0.
\ee

For an asymptotically AdS spacetime, this means setting the dynamical exponent to $z=1$, gives us a very interesting result. Namely, eq(\ref{nfl_cond}) gives the conductivity of the Fermi liquid (FL) whereas eq(\ref{fl_cond}) gives the conductivity of the non-Fermi liquid (NFL). So, it follows that by doing appropriate fine tuning of the magnetic field and the charge density, we can see the existence of either the FL phase or the NFL phase. Hence, the master formula for the conductivity eq(\ref{master_formula_cond}) describes the unification of  both the FL and NFL phase.

\subsection{Rotationally symmetric spacetime }
In this subsection, we shall consider a rotationally symmetric spacetime, $g_{xx}=g_{yy}=g_{zz}$ in
$4+1$-dimensional bulk spacetime and show the absence of the exact  equality among all the diagonal components of the conductivity tensor and reproduce the result of the previous section using the RG flow of the conductivity technique as mentioned in the introduction. It means, it is not necessary that all the diagonal components of the conductivity tensor should be same, i.e. $\sigma_{ii}\neq \sigma_{jj}$, where there is no sum over the repeated index, both $i$ and $j$ run over the spatial dimensions. The easiest way to see such a behavior is by turning on a magnetic field, say along the x-direction, in the probe brane approximation. In which case, it follows that  $\sigma^{xx}\neq \sigma^{yy}=\sigma^{zz}$. This we demonstrate by adopting the RG flow of the conductivity for the DBI type of action   as in \cite{ssp1} but in a slightly different setting. We show that the dc conductivity calculated using the approach of \cite{kob} can be reproduced to leading order in the electric field using the technique of \cite{il} for the DBI action as in \cite{ssp1}.

\subsubsection{In-equality of the diagonal conductivity}

The induced metric and the two form field strength with fluctuation that we consider has the following form
\bea
ds^2(ind)&=&-g_{tt}(r)dt^2+g_{xx}(r)dx^2+g_{yy}(r)dy^2+g_{zz}(r)dz^2+g_{rr}(r)dr^2\nn
F_2&=&A'_t(r)dr\w dt+Bdy\w dz\nn&+&F^{(1)}_{tx^i}dt\w dx^i+F^{(1)}_{tr}dt\w dr+F^{(1)}_{rx^i}dr\w dx^i+F^{(1)}_{x^ix^j}dx^i\w dx^j,
\eea
where we have written the fluctuations in the superscript of the fields as $(1)$ and allow the fluctuations to depend on $(t,~x,~y,~z,~r)$. We denote  $(x^1\equiv x,~x^2\equiv y,~x^3\equiv z)$.  The equation of motion of the  fluctuating gauge field 
take the following form
\bea
&&\f{\p_tF^{(1)}_{rt}g_{tt}g_{rr}}{(g_{tt}g_{rr}-A'^2_t)^2}+\f{g_{tt}\p_x F^{(1)}_{xr}}{g_{xx}(g_{tt}g_{rr}-A'^2_t)}+\f{(g_{tt}g_{zz}\p_yF^{(1)}_{yr}+
g_{tt}g_{yy}\p_zF^{(1)}_{zr})}{(g_{tt}g_{rr}-A'^2_t)(g_{yy}g_{zz}+B^2)}=0\nn
&&\p_r\Bigg[\f{\sqrt{g_{xx}(g_{yy}g_{zz}+B^2)}}{(g_{tt}g_{rr}-A'^2_t)^{3/2}}g_{tt}g_{rr}F^{(1)}_{tr}+\f{BA'_t\sqrt{g_{xx}}
F^{(1)}_{zy}}{\sqrt{g_{tt}g_{rr}-A'^2_t}\sqrt{g_{yy}g_{zz}+B^2}} \Bigg]-\nn
&&\f{\sqrt{g_{xx}}[g_{zz}g_{rr}\p_yF^{(1)}_{yt}+ g_{yy}g_{rr}\p_zF^{(1)}_{zt}-BA'_t\p_rF^{(1)}_{yz}]}{\sqrt{g_{tt}g_{rr}-A'^2_t}\sqrt{g_{yy}g_{zz}+B^2}}-
\f{g_{rr}\sqrt{g_{yy}g_{zz}+B^2}}{\sqrt{g_{xx}(g_{tt}g_{rr}-A'^2_t)}}\p_xF^{(1)}_{xt}=0\nn
&&\p_r\Bigg[\f{g_{tt}\sqrt{g_{yy}g_{zz}+B^2}~F^{(1)}_{rx}}{\sqrt{g_{xx}(g_{tt}g_{rr}-A'^2_t)}} \Bigg]-\f{g_{rr}\sqrt{g_{yy}g_{zz}+B^2}~\p_tF^{(1)}_{tx}}{\sqrt{g_{xx}(g_{tt}g_{rr}-A'^2_t)}}+\f{\sqrt{g_{tt}g_{rr}-A'^2_t}[g_{zz}\p_y F^{(1)}_{yx}+g_{yy}\p_z F^{(1)}_{zx}]}{\sqrt{g_{xx}(g_{yy}g_{zz}+B^2)}}=0\nn
&&\p_r\Bigg[\f{\sqrt{g_{xx}}(BA'_tF^{(1)}_{tz}+g_{tt}g_{zz}F^{(1)}_{ry})}{\sqrt{g_{tt}g_{rr}-A'^2_t}\sqrt{g_{yy}g_{zz}+B^2}} \Bigg]+\f{\sqrt{g_{xx}}[BA'_t\p_rF^{(1)}_{zt}-g_{rr}g_{zz}\p_tF^{(1)}_{ty}]}{\sqrt{g_{tt}g_{rr}-A'^2_t}\sqrt{g_{yy}g_{zz}+B^2}}\nn&+&\f{\sqrt{g_{tt}g_{rr}-A'^2_t}g_{zz}\p_x F^{(1)}_{xy}}{\sqrt{g_{xx}(g_{yy}g_{zz}+B^2)}}+\f{\sqrt{g_{xx}(g_{tt}g_{rr}-A'^2_t)}g_{yy}g_{zz}\p_z F^{(1)}_{zy}}{(g_{yy}g_{zz}+B^2)^{3/2}}=0\nn
&&\p_r\Bigg[\f{\sqrt{g_{xx}}(g_{tt}g_{yy}F^{(1)}_{rz}-BA'_tF^{(1)}_{ty})}{\sqrt{g_{tt}g_{rr}-A'^2_t}\sqrt{g_{yy}g_{zz}+B^2}} \Bigg]+\f{\sqrt{g_{xx}}[BA'_t\p_rF^{(1)}_{ty}-g_{rr}g_{yy}\p_tF^{(1)}_{tz}]}{\sqrt{g_{tt}g_{rr}-A'^2_t}\sqrt{g_{yy}g_{zz}+B^2}}\nn&+&\f{\sqrt{g_{tt}g_{rr}-A'^2_t}g_{yy}\p_x F^{(1)}_{xz}}{\sqrt{g_{xx}(g_{yy}g_{zz}+B^2)}}+\f{\sqrt{g_{xx}(g_{tt}g_{rr}-A'^2_t)}g_{yy}g_{zz}\p_y F^{(1)}_{yz}}{(g_{yy}g_{zz}+B^2)^{3/2}}=0.
\eea

The first equation is essentially a constraint. The expression of the currents at a constant slice of $r$ are
\bea\label{current_5d}
J^{(1)x}&=&- \f{T_b\sqrt{g_{yy}g_{zz}+B^2}}{\sqrt{g_{xx}(g_{tt}g_{rr}-A'^2_t)}}g_{tt}F^{(1)}_{rx},\quad
J^{(1)y}=-\f{T_b\sqrt{g_{xx}}[g_{tt}g_{zz}F^{(1)}_{ry}+BA'_t F^{(1)}_{tz}]}{\sqrt{g_{tt}g_{rr}-A'^2_t}\sqrt{g_{yy}g_{zz}+B^2}},\nn
J^{(1)z}&=&-T_b\f{\sqrt{g_{xx}}}{\sqrt{g_{tt}g_{rr}-A'^2_t}\sqrt{g_{yy}g_{zz}+B^2}}[g_{tt}g_{yy}F^{(1)}_{rz}-BA'_t F^{(1)}_{ty}].
\eea

Hence forth, we shall work in a gauge choice: $A_r=0$ and the Fourier transformation will be done with respect to $e^{i(k_x x+k_yy+k_zz-\omega t)}$. In which case, the equation of motion becomes
\bea\label{all_eom_gauge_fixed_5d}
&&\omega g_{xx}g_{rr}(g_{yy}g_{zz}+B^2)\p_r A^{(1)}_{t}+\nn&& (g_{tt}g_{rr}-A'^2)[k_x(g_{yy}g_{zz}+B^2)\p_r A^{(1)}_{x}+k_yg_{xx}g_{zz}\p_r A^{(1)}_{y}+k_zg_{xx}g_{yy}\p_r A^{(1)}_{z}]=0\nn
&&\p_r\Bigg[\f{\sqrt{g_{xx}}[(g_{yy}g_{zz}+B^2)g_{tt}g_{rr}\p_r A^{(1)}_{t}-iBA'_t(g_{tt}g_{rr}-A'^2_t)E^{(1)}_{zy}]}{(g_{tt}g_{rr}-A'^2_t)^{3/2}
\sqrt{g_{yy}g_{zz}+B^2}}\Bigg]-
\f{g_{rr}\sqrt{g_{yy}g_{zz}+B^2}}{\sqrt{g_{xx}(g_{tt}g_{rr}-A'^2_t)}}k_xE^{(1)}_x-\nn
&&\f{\sqrt{g_{xx}}[k_y g_{zz}g_{rr} E^{(1)}_y+k_zg_{yy}g_{rr}E^{(1)}_z+iBA'_t\p_rE^{(1)}_{yz}]}{\sqrt{g_{tt}g_{rr}-A'^2_t}\sqrt{g_{yy}g_{zz}+B^2}}=0\nn
&&\p_r\Bigg[\f{g_{tt}\sqrt{g_{yy}g_{zz}+B^2}\p_r A^{(1)}_{x}}{\sqrt{g_{xx}(g_{tt}g_{rr}-A'^2_t)}} \Bigg]+\f{g_{rr}\sqrt{g_{yy}g_{zz}+B^2}\omega E^{(1)}_x}{\sqrt{g_{xx}(g_{tt}g_{rr}-A'^2_t)}}-\f{\sqrt{g_{tt}g_{rr}-A'^2_t}[g_{zz}k_yE^{(1)}_{yx}+g_{yy}k_z E^{(1)}_{zx}]}{\sqrt{g_{xx}(g_{yy}g_{zz}+B^2)}}=0\nn
&&\p_r\Bigg[\f{\sqrt{g_{xx}}(g_{tt}g_{zz}\p_rA^{(1)}_{y}-iBA'_t E^{(1)}_z)}{\sqrt{g_{tt}g_{rr}-A'^2_t}\sqrt{g_{yy}g_{zz}+B^2}}\Bigg]+\f{\sqrt{g_{xx}}[\omega g_{rr}g_{zz}E^{(1)}_y+iBA'_t\p_r E^{(1)}_z]}{\sqrt{g_{tt}g_{rr}-A'^2_t}\sqrt{g_{yy}g_{zz}+B^2}}-\nn
&&\f{\sqrt{g_{tt}g_{rr}-A'^2_t}}{\sqrt{g_{xx}}(g_{yy}g_{zz}+B^2)^{3/2}}\bigg[k_x(g_{yy}g_{zz}+B^2)g_{zz}E^{(1)}_{xy}+k_z g_{xx}g_{yy}g_{zz}E^{(1)}_{zy}\bigg]=0\nn
&&\p_r\Bigg[\f{\sqrt{g_{xx}}(g_{tt}g_{yy}\p_rA^{(1)}_{z}+iBA'_t E^{(1)}_y)}{\sqrt{g_{tt}g_{rr}-A'^2_t}\sqrt{g_{yy}g_{zz}+B^2}}\Bigg]+\f{\sqrt{g_{xx}}[\omega g_{rr}g_{yy}E^{(1)}_z-iBA'_t\p_r E^{(1)}_y]}{\sqrt{g_{tt}g_{rr}-A'^2_t}\sqrt{g_{yy}g_{zz}+B^2}}-\nn
&&\f{\sqrt{g_{tt}g_{rr}-A'^2_t}}{\sqrt{g_{xx}}(g_{yy}g_{zz}+B^2)^{3/2}}
\bigg[k_x(g_{yy}g_{zz}+B^2)g_{yy}E^{(1)}_{xz}+k_y g_{xx}g_{yy}g_{zz}E^{(1)}_{yz}\bigg]=0,
\eea
where we have defined the quantities, $E^{(1)}_i\equiv \omega A^{(1)}_i+k_i A^{(1)}_t$ and $E^{(1)}_{ij}\equiv k_i A^{(1)}_j-k_j A^{(1)}_i$. Here $i$ and $j$ can take values $x,~y$ and $z$. Generically, it is very difficult to find the decoupled equation of motions for each component of the gauge field. So we shall make a specific choice to the value of the  momenta, in which case, things become simple to find the  decoupled equation of motions.\\

Choice: $k_y=0=k_z$\\

For this particular choice of the momenta, the linear combination of the gauge field component $A^{(1)}_x$ and
$A^{(1)}_t$, which is $E^{(1)}_x$ decouples from the other equations and it takes the following form
\be
\p_r\Bigg[\f{g_{tt}g_{rr}\sqrt{g_{xx}(g_{yy}g_{zz}+B^2)}\p_r E^{(1)}_x}{\sqrt{g_{tt}g_{rr}-A'^2_t}[\omega^2g_{xx}g_{rr}-k^2_x(g_{tt}g_{rr}-A'^2_t)]} \Bigg]+\f{g_{rr}\sqrt{g_{yy}g_{zz}+B^2}}{\sqrt{g_{xx}(g_{tt}g_{rr}-A'^2_t)}}E^{(1)}_x=0.
\ee

Differentiating the Ohm's law,  $J^{(1)x}=\sigma^{xx}
F^{(1)}_{xt}=i\sigma^{xx}
E^{(1)}_x$, with respect to $r$  gives us the flow equation for the conductivity 
\bea\label{flow_sigma_xx_5d_kx}
&&\p_r\sigma^{xx}=\f{i\omega \sqrt{g_{rr}}}{\sqrt{g_{tt}}g_{xx}T_b[\rho^2+g_{xx}(g_{yy}g_{zz}+B^2)]^{3/2}}\times\nn&&\Bigg[(\sigma^{xx})^2 g^2_{xx}\bigg( \rho^2 +(g_{yy}g_{zz}+B^2)(g_{xx}-\f{k^2_x}{\omega^2}g_{tt})\bigg)-T^2_b\bigg(\rho^2+g_{xx}(g_{yy}g_{zz}+B^2) \bigg)^2\Bigg],
\eea
where $\rho$ is the charge density and also  the solution to $A'_t$
\be
A'_t(r)=\f{\rho\sqrt{g_{tt}g_{rr}}}{\sqrt{\rho^2+g_{xx}(g_{yy}g_{zz}+B^2)}},\quad g_{tt}g_{rr}-A'^2_t=\f{g_{tt}g_{rr}g_{xx}(g_{yy}g_{zz}+B^2)}{\rho^2+g_{xx}(g_{yy}g_{zz}+B^2)}.
\ee

The regularity condition of the conductivity at the horizon, $r_h$, gives the following condition
\be\label{cond_horizon_x_5d}
Re\sigma^{xx}(r_h)=T_b\Bigg[\f{\sqrt{\rho^2+g_{xx}(g_{yy}g_{zz}+B^2)}}{g_{xx}}\Bigg]_{r_h}.
\ee

It is very easy to see that this result of the conductivity precisely matches with the  leading order  result in the electric field as written in  eq(\ref{special_long_cond}).

The other components of the gauge field $A^{(1)}_y$ and $A^{(1)}_z$ can be combined together as $A^{(1)}_{\pm}\equiv A^{(1)}_y\pm i A^{(1)}_z$, in which case the equation of motion of $A^{(1)}_+$ decouples from the $A^{(1)}_-$.
The explicit form of the equation of motions reads as
\bea
&&\p_r\Bigg[\f{\sqrt{g_{xx}}(g_{tt}g_{yy}\p_r A^{(1)}_{\pm}\mp\omega BA'_t A^{(1)}_{\pm})}{\sqrt{g_{tt}g_{rr}-A'^2_t}\sqrt{g_{yy}g_{zz}+B^2}} \Bigg]+\f{\omega\sqrt{g_{xx}}(\omega g_{rr}g_{yy}A^{(1)}_{\pm}\pm BA'_t \p_r A^{(1)}_{\pm})}{\sqrt{g_{tt}g_{rr}-A'^2_t}\sqrt{g_{yy}g_{zz}+B^2}}\nn
&&-k^2_x\f{g_{yy}\sqrt{g_{tt}g_{rr}-A'^2_t}}{\sqrt{g_{xx}(g_{yy}g_{zz}+B^2)}}A^{(1)}_{\pm}=0.
\eea

Defining the currents at a constant slice of $r$ as
\be
J^{(1)}_{\pm}\equiv J^{(1)}_y\pm i J^{(1)}_z=-\f{T_b\sqrt{g_{xx}}}{\sqrt{g_{tt}g_{rr}-A'^2_t}\sqrt{g_{yy}g_{zz}+B^2}}\Bigg[g_{tt}g_{yy}\p_r A^{(1)}_{\pm}\mp i\omega B A'_t A^{(1)}_{\pm} \Bigg]
\ee

and the conductivity as $\sigma_{\pm}\equiv\sigma^{yz}\pm i\sigma^{yy}$, where we have assumed $\sigma^{yy}=\sigma^{zz}$ and $\sigma^{zy}=-\sigma^{yz}$  for $g_{yy}=g_{zz}$. In this case, the Ohm's law becomes
\be\label{ohms_law_5d}
J^{(1)}_{\pm}=\pm\omega \sigma_{\pm} A^{(1)}_{\pm},
\ee
and the conductivity, $\sigma_{\pm}$, is related to $A^{(1)}_{\pm}$ as
\be\label{def_sigma_pm_5d}
\sigma_{\pm}=\pm \f{T_b\sqrt{g_{xx}}}{\sqrt{g_{tt}g_{rr}-A'^2_t}\sqrt{g_{yy}g_{zz}+B^2}}\Bigg[-\f{g_{tt}g_{yy}\p_r A^{(1)}_{\pm}}{\omega A^{(1)}_{\pm}}\pm B A'_t\Bigg].
\ee
As we go from one slice of constant $r$ to another   there follows the flow equation of the conductivity, upon using eq(\ref{ohms_law_5d}) and eq(\ref{def_sigma_pm_5d})
\bea
\p_r\sigma_{\pm}&=&\pm\f{T_b[k^2_x g_{tt}g^2_{yy}(g_{tt}g_{rr}-A'^2_t)+\omega^2 g_{xx}(g_{tt}g_{rr}g^2_{yy}+B^2A'^2_t)]}{\omega g_{tt}g_{yy}\sqrt{g_{xx}(g_{tt}g_{rr}-A'^2_t)}\sqrt{g_{yy}g_{zz}+B^2}}\mp \f{2\omega B A'_t\sigma_{\pm}}{g_{tt}g_{yy}}\nn &\pm& \f{\omega\sigma^2_{\pm}\sqrt{g_{tt}g_{rr}-A'^2_t}\sqrt{g_{yy}g_{zz}+B^2}}{T_b \sqrt{g_{xx}}g_{tt}g_{yy} }.
\eea

Using the solution to, $A'_t$, we end up with the following form of the flow equation for the conductivity
\bea
\p_r\sigma^{yz}&=&\f{i2\omega\sigma^{yy}\sqrt{g_{rr}}}{T_b g_{yy}\sqrt{g_{tt}[\rho^2+g_{xx}(g_{yy}g_{zz}+B^2)]}}
\Bigg[\sigma^{yz}(g_{yy}g_{zz}+B^2)-T_b \rho B \Bigg]\nn
\p_r\sigma^{yy}&=&-\f{i\omega\sqrt{g_{rr}} \Bigg[T^2_b[\rho^2+g_{xx}g^2_{yy}+
(\f{k^2_x}{\omega^2})g_{tt}g^2_{yy}]+(g^2_{yy}+B^2)((\sigma^{yz})^2-(\sigma^{yy})^2)-2T_b\rho B\sigma^{yz} \Bigg]}{T_b g_{yy}\sqrt{g_{tt}[\rho^2+g_{xx}(g_{yy}g_{zz}+B^2)]}}\nn
\eea

The regularity condition on the conductivity at the horizon, which is same as imposing the in-falling boundary condition for the gauge field at the horizon \cite{ssp1}, gives
\bea\label{cond_horizon_y_z_5d}
Re\sigma^{yz}(r_h)&=&T_b\f{\rho B}{g^2_{yy}(r_h)+B^2},\quad Im\sigma^{yz}(r_h)=0,\nn
Re\sigma^{yy}(r_h)&=&\pm T_b\Bigg[\f{g_{yy}\sqrt{\rho^2+g_{xx}(g^2_{yy}+B^2)}}{g^2_{yy}+B^2}\Bigg]_{r_h},\quad Im\sigma^{yy}(r_h)=0
\eea

From the regularity condition of the conductivity at the  horizon  eq(\ref{cond_horizon_x_5d}) and eq(\ref{cond_horizon_y_z_5d}) there follows
\bea
\f{Re\sigma^{yy}(r_h)}{Re\sigma^{yz}(r_h)}&=&\pm \f{g_{yy}\sqrt{\rho^2+g_{xx}(g^2_{yy}+B^2)}}{\rho B},\quad \f{Re\sigma^{yy}(r_h)}{Re\sigma^{xx}(r_h)}=\pm \f{g_{yy}g_{xx}}{g^2_{yy}+B^2},\nn
\f{Re\sigma^{xx}(r_h)}{Re\sigma^{yz}(r_h)}&=&\f{\sqrt{\rho^2+g_{xx}(g^2_{yy}+B^2)}[g^2_{yy}+B^2]}{g_{xx}\rho B}.
\eea

If we consider the spacetime to be of the following form $g_{xx}=g_{yy}=g_{zz}$ then the diagonal components of the conductivity tensor are not same.  In order to have a feel for the temperature dependence in the calculations, let us consider the spacetime to be of the Lifshitz type where $g_{xx}=g_{yy}=g_{zz}=r^2$ and using $r_h\sim T^{1/z}$, where $T$ is the Hawking temperature and $z$ is the dynamical exponent (which should not be confused with the spatial coordinate $z$). In  the very high charge density and low magnetic field limit
\be
Re\sigma^{xx}\sim Re\sigma^{yy} \sim \rho T^{-2/z},\quad Re\sigma^{yz}\sim \rho B T^{-4/z}.
\ee

In the low charge density and high magnetic field limit
\be
Re\sigma^{xx}\sim  B T^{-1/z},\quad Re\sigma^{yz}={\rm Constant},\quad Re\sigma^{yy} \sim  T^{3/z}/B.
\ee

For unit dynamical exponent, $z=1$, as we tune the charge density and the magnetic field the conductivity, $\sigma^{xx}$ go from non-Fermi liquid phase to Fermi liquid phase. Moreover, in the high charge density and low magnetic field the ratio $Re\sigma^{yy}/Re\sigma^{yz}\sim T^2$, which is reminiscent of the behavior of the inverse Hall angle of NFL liquid.  The  behavior of the conductivity that matches with the experimental result is presented in the table below.

\begin{tabular}{ | l |l| l | l |l|}
    \hline
 Type & Choice of $ z$ &Physical quantity & Limit &Field theory \\
       &  &at the horizon   &        &     dimension\\
\hline\hline
  & && Low charge density $\&$& \\
NFL     &$z=1$      & Re $\sigma^{xx}\sim T^{-1}$                          & High magnetic field&$3+1$ dimension\\
 \hline\hline
 & && High charge density $\&$& \\
FL     &$z=1$      & Re $\sigma^{xx}\sim T^{-2}$                          & Low magnetic field&$3+1$ dimension\\
 \hline\hline
    \end{tabular}
\begin{center}
\noindent 
Table 2: The longitudinal conductivity that matches with the experimental result for unit dynamical exponent.
\end{center}
However, for unit dynamical exponent the inverse Hall angle  does not matches with the experimental result for the non-Fermi liquid state. This is summarized below in a table.

\begin{tabular}{ | l |l| l | l |l|}
    \hline
 Type & Choice of $ z$ &Physical quantity & Limit &Field theory \\
       &  &at the horizon   &        &     dimension\\
\hline\hline
  & &$\f{{\rm Re} \sigma^{xx}}{{\rm Re}\sigma^{yz}}\sim T^{-1}$ & Low charge density $\&$& \\
NFL ?    &$z=1$      &                          & High magnetic field&$3+1$ dimension\\
& & $\f{{\rm Re} \sigma^{yy}}{{\rm Re}\sigma^{yz}}\sim T^{3}$ & &\\
 \hline\hline
 & &$\f{{\rm Re} \sigma^{xx}}{{\rm Re}\sigma^{yz}}\sim T^{2}$ & High charge density $\&$& \\
FL  ?   &$z=1$      &                          & Low magnetic field&$3+1$ dimension\\
& & $\f{{\rm Re} \sigma^{yy}}{{\rm Re}\sigma^{yz}}\sim T^{2}$ & &\\
 \hline\hline
    \end{tabular}
\begin{center}
\noindent 
Table 3: The longitudinal conductivity that does not matches with the experimental result for unit dynamical exponent.
\end{center}

\subsection{Some consequences}

In this subsection, we shall analyze the consequences of the behavior of the conductivity at the horizon. Let us write them down explicitly in terms of the temperature
\bea
Re\sigma^{xx}(r_h)&=&T_b~ T^{-2/z}\sqrt{\rho^2+T^{6/z}+B^2 T^{2/z}},\quad
Re\sigma^{yz}(r_h)=T_b\f{\rho B}{B^2+T^{4/z}},\nn Re\sigma^{yy}(r_h)&=&Re\sigma^{zz}(r_h)=T_b~ T^{2/z}\f{\sqrt{\rho^2+T^{6/z}+B^2 T^{2/z}}}{B^2+T^{4/z}},
\eea
where we have used the relation between the size of the horizon and the black hole temperature as $r_h=T^{1/z}$. The imaginary part of the conductivities vanishes.
Generically, it is very difficult to find the precise location of the temperature where there occurs a minimum of the conductivity, analytically.

From the expression of the $\sigma^{xx}$ conductivity, it follows that at zero temperature, $T\ra 0$, the conductivity diverges, it means the resistivity drops to zero. So it suggests as we approach zero temperature the system has nearly zero resistance.  Similarly, from the expression of the  conductivity, $\sigma^{yy}$, for $z=1$  at the horizon, it follows that at zero temperature, $T\ra 0$, the conductivity vanishes, it means the resistivity diverges. So, the behavior of the conductivity along the direction of the magnetic field and perpendicular to it are completely different.

\begin{figure}[htb]
\centering
\subfigure[1/Re$\sigma^{xx}$ versus $T$]
{\includegraphics[width=3.0 in]
{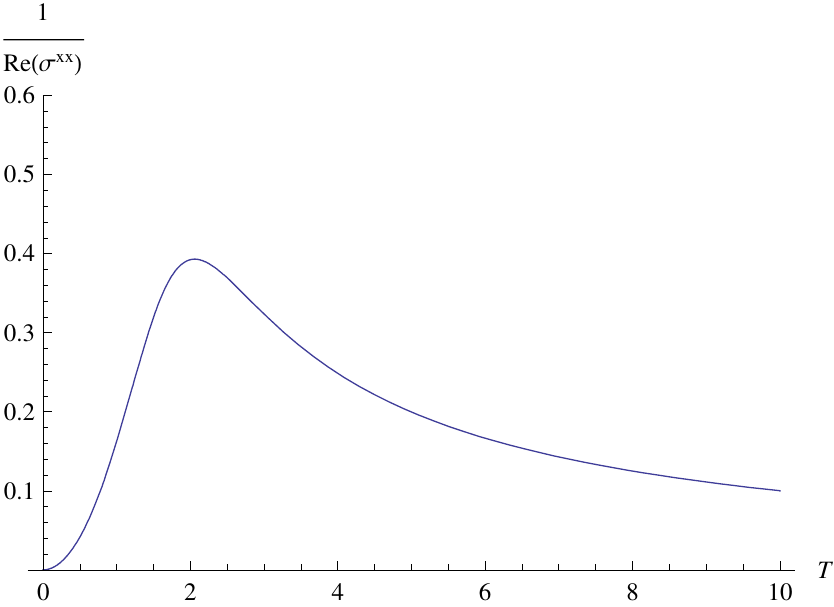}}
\subfigure[1/Re$\sigma^{yy}$ versus $T$]
{\includegraphics[width=3.0 in]
{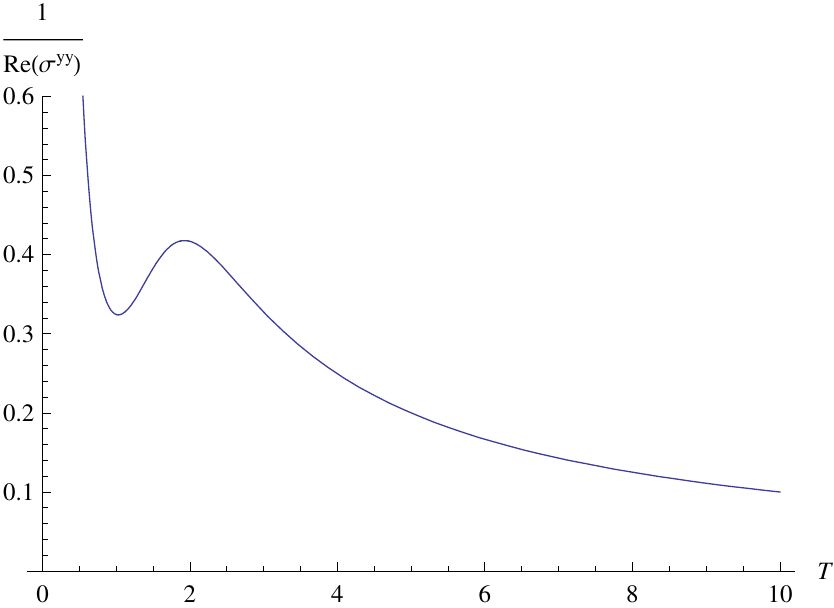}}
\caption{The  behavior of the
DC resistivity for unit dynamical exponent and unit tension of the brane with $\rho=6$ and $B=1$.}
\label{fig_3}
\end{figure}

\begin{figure}[htb]
\centering
\subfigure[1/Re$\sigma^{xx}$]
{\includegraphics[width=3.0 in]
{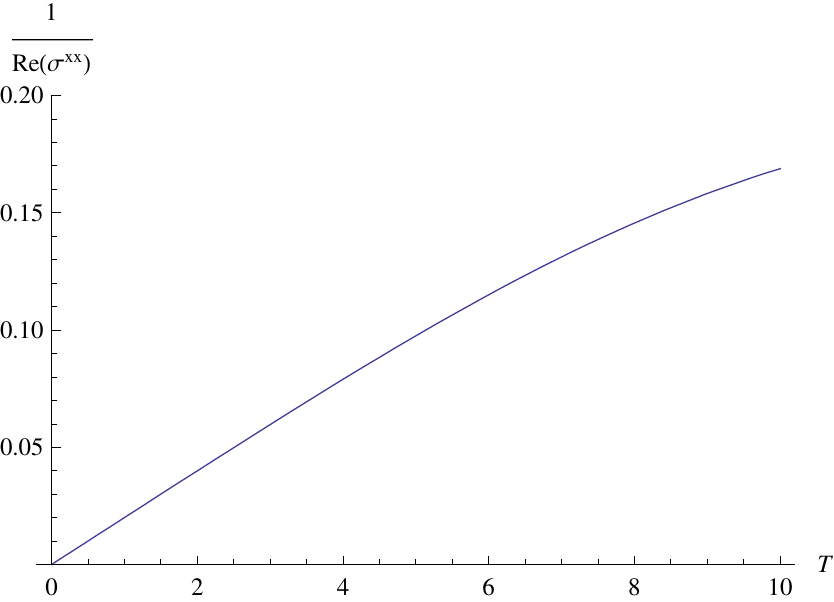}}
\subfigure[1/Re$\sigma^{yy}$]
{\includegraphics[width=3.0 in]
{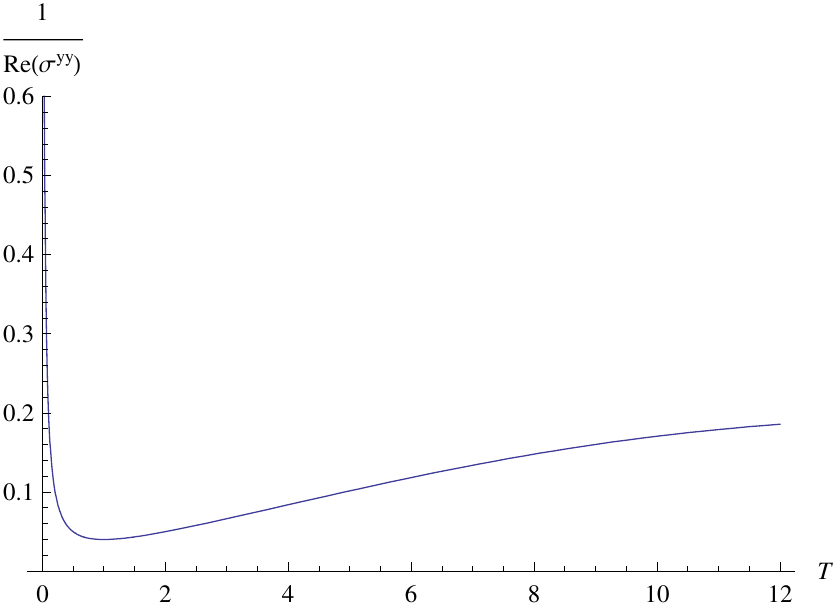}}
\caption{The  behavior of the DC resistivity for $z=2$  and unit tension of the brane with $\rho=50$ and $B=1$.}
\label{fig_4}
\end{figure}

The behavior of the conductivities as a function of the temperature  for fixed charge density and magnetic field are plotted in fig(\ref{fig_3}) and fig(\ref{fig_4}) for different choice of the dynamical exponent.
Even though both the resistivities,  $1/Re\sigma^{xx}$ and $1/Re\sigma^{yy}$ show the linear temperature dependence at high charge density and low magnetic field limit, the important upturn behavior is only seen for the  resistivity along the direction perpendicular to the magnetic field, $1/Re\sigma^{yy}$.
From the expression of the dc Hall conductivity, it follows that  the Hall coefficient  has a  power law behavior at high temperature for non-zero charge density and magnetic field, $R_H\sim T^{4/z}$.

\begin{figure}[htb]
\centering
\subfigure[Re$\sigma^{yz}$]
{\includegraphics[width=3.0 in]
{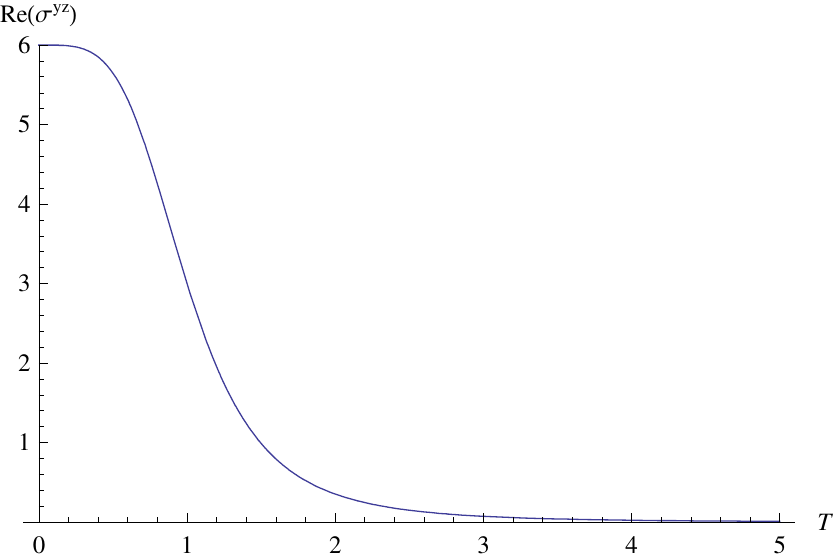}}
\subfigure[$R_H~\rho$]
{\includegraphics[width=3.0 in]
{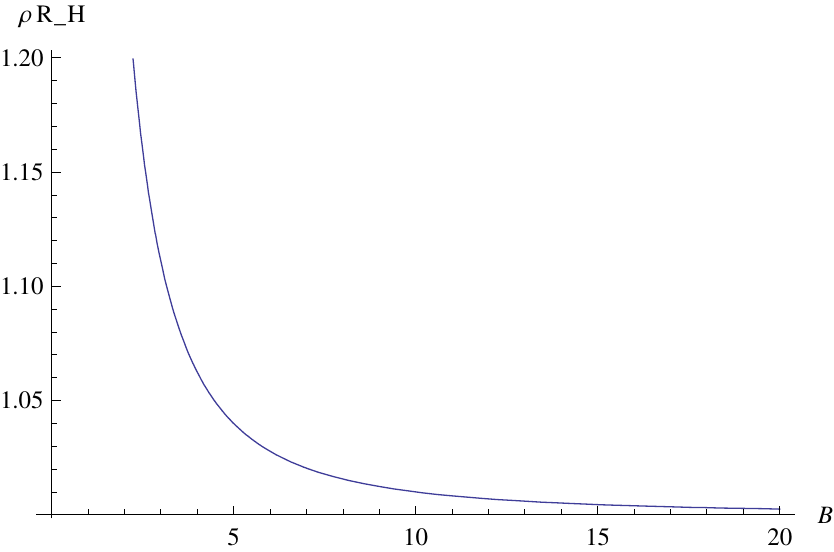}}
\caption{The  behavior of the DC Hall  conductivity for unit dynamical exponent and unit tension of the brane with $\rho=6$ and $B=1$ in (a) and in (b) the Hall coefficient, $R_H=(\sigma^{yz}B)^{-1}$, times the charge density is plotted  versus magnetic field for unit temperature.}
\label{fig_5}
\end{figure}

\subsection{Diffusion constant}
The diffusion constant can very easily be derived from the  flow equation of the conductivity in the hydrodynamic regime, $\omega \sim k^2_x\ra 0,~\omega \ll T$ and $k_x\ll T$. In which case, the flow equation of the conductivity approximates to
\be
\p_r\sigma^{xx}\simeq -i\f{k^2_x}{\omega}\f{\sqrt{g_{rr}g_{tt}}(\sigma^{xx})^2g_{xx}(g_{yy}g_{zz}+B^2)}{T_b[\rho^2+g_{xx}(g_{yy}g_{zz}+B^2)]^{3/2}}.
\ee
Integrating out the flow equation for the conductivity from the horizon, $r_h$, to the boundary, $r=\infty$ and using the relation between the Green's function  and the conductivity at the boundary
\be
G^{xx}_R(\omega,~k_x)=-i\omega \sigma^{xx}(r\ra\infty)=
\f{\omega^2\sigma^{xx}(r_h)}{i\omega-Dk^2_x},
\ee
gives the diffusion constant as
\be
D=\f{\sigma^{xx}(r_h)}{T_b}\int^{\infty}_{r_h}
\f{\sqrt{g_{rr}g_{tt}}g_{xx}(g_{yy}g_{zz}+B^2)}{[\rho^2+g_{xx}(g_{yy}g_{zz}+B^2)]^{3/2}}.
\ee

\subsection{Magnetic conductivity}
The use of the Ohm's law for the  electrical conductivity, $J^i=\sigma^{ij}F_{jt}$, where $F_{jt}$ is the field strength and $i,~j$ runs over the spatial coordinates, helps us to determine the flow equation for the conductivity. Similarly, we can determine the magnetic conductivity via the formula, $J^i=\sigma^{ij}_mB_j=
\sigma^{ij}_m\epsilon_{jlm}F_{lm}$, where $B_j$ is the magnetic field and $\epsilon_{ijk}$ is the Levi-Civita tensor, which can take values $\pm 1$ and $0$. For example, with three spatial coordinates,  $x,~y,~z$, in the Fourier space, the current
\be
J^z=-2i[(\sigma^{zx}_mk_y-
\sigma^{zy}_mk_x)A_z+
(\sigma^{zz}_mk_x-
\sigma^{zx}_mk_z)A_y+
(\sigma^{zy}_mk_z-
\sigma^{zz}_mk_y)A_x],
\ee
where $F_{ij}=\p_iA_j-\p_j A_i$ and $k_i$ is the momentum along the i'th direction with the convention $\epsilon_{xyz}=-1$. For vanishing $A_x=0,~A_y=0$ and the momenta $k_y=0=k_z$, the current reduces to $J^z=2ik_x \sigma^{zy}_mA_z$. Let us look at the equations of motion as written in  eq(\ref{all_eom_gauge_fixed_5d})
for zero frequency and with  $A^{(1)}_x=0=A^{(1)}_y$, in which case, the equations of motion of $A^{(1)}_t$ and $A^{(1)}_z$  decouple. The current, $J^{(1)z}$, at a constant slice of $r$ and the equation of motion of $A^{(1)}_z$ are
\bea
&&J^{(1)z}=-T_b\f{\sqrt{g_{xx}}g_{tt}g_{yy}\p_rA^{(1)}_{z}}{\sqrt{g_{tt}g_{rr}-A'^2_t}\sqrt{g_{yy}g_{zz}+B^2}}\nn
&&\p_r\Bigg[\f{\sqrt{g_{xx}}g_{tt}g_{yy}\p_rA^{(1)}_{z}}{\sqrt{g_{tt}g_{rr}-A'^2_t}\sqrt{g_{yy}g_{zz}+B^2}}\Bigg]
-\f{g_{yy}\sqrt{g_{tt}g_{rr}-A'^2_t}}{\sqrt{g_{xx}(g_{yy}g_{zz}+B^2)}}k^2_x A^{(1)}_z=0
\eea

Using the previous prescription and the solution of $A'_t$, we find the following flow equation for the  magnetic conductivity
\be
\p_r\sigma^{zy}_m=
\f{ik_x\sqrt{g_{rr}}[T^2_b g_{tt}g^2_{yy}+
4(\sigma^{zy})^2(g^2_{yy}+B^2)]}{2T_b g_{yy}\sqrt{g_{tt}}
\sqrt{\rho^2+g_{xx}(g^2_{yy}+B^2)}}. 
\ee
The regularity requires us to set the following condition at the horizon, $\sigma^{zy}_m(r_h)=0$. Similarly, we can find the flow equation at non-zero frequency, $\omega\neq 0$. In order for the gauge field component $A^{(1)}_z$ to decouple from the  other components, we need to set   $B=0$ along with $A^{(1)}_x=0=A^{(1)}_y$ and $k_y=0=k_z$ in eq(\ref{all_eom_gauge_fixed_5d}).
In which case, the flow equation becomes
\be
\p_r\sigma_m^{zy}=-
\f{i\sqrt{g_{rr}}}{2k_x T_b g_{yy}\sqrt{g_{tt}}
\sqrt{\rho^2+g_{xx}g^2_{yy}}}
\Bigg[T^2_b\bigg(\omega^2
(\rho^2+g_{xx}g^2_{yy})-k^2_x
g_{tt}g^2_{yy}\bigg)-4k^2_x g^2_{yy}(\sigma^{zy})^2\Bigg].
\ee
The regularity condition at the horizon, $r_h$, requires us to set
\be
\sigma^{zy}_m(r_h)=T_b\Bigg[\f{\omega
\sqrt{\rho^2+g_{xx}g^2_{yy}}}{2k_xg_{yy}}\Bigg]_{r_h},
\ee
which in the hydrodynamic limit, $\omega\sim k^2_x\ra 0$, implies the vanishing of the conductivity at the horizon, $\sigma^{zy}_m(r_h)\ra 0$. Note that,  in contrast to the electrical conductivity at the horizon, the magnetic conductivity at the horizon depends on the frequency and the momentum. 

\subsection{Flow equation with DBI and Chern-Simon term}

In this subsection, we shall find the change to the flow equation for the conductivity upon the inclusion of the Chern-Simon term to the action. Let us assume that the form of the action is same as in  string theory
\be
S_{CS}=\mu_4\int\Bigg[ C_5+C_3\w F_2+\f{C_1}{2}\w F_2\w F_2 \Bigg],
\ee
where $C_n$ are the RR n-form potential. With the field strength
$F^{(0)}_2=A'_t (r)dr\w dt+B dy\w dz$, the solution to the equation of motion that results from both the DBI and CS action in the massless limit i.e. for trivial embeddings
\be
A'^2_t=\f{g_{tt}g_{rr}(\rho-2 B c^{(1)}_{x})^2}{g_{xx}(g^2_{yy}+B^2)+(\rho-2 B c^{(1)}_{x})^2},
\ee
where $\rho$ is the charge density and we have considered the dilaton, $\phi=0$, to be trivial,  $B_2=0,~C_3=C_{trx}(r)dt\w dr\w dx$ and $C_1=c^{(1)}_{x}(r) dx$.
Upon doing the fluctuation to the gauge potential, $A_M\ra A_M+A^{(1)}_M(t,r,x)$,  the  quadratically   fluctuated action is
\bea
S^{(2)}&=&-T_4\int \sqrt{-det M^{(0)}_+}\bigg[-
\f{1}{4}{\bigg(M^{(0)}_+\bigg)^{-1}}^{ab}F^{(1)}_{bc}{\bigg(M^{(0)}_+\bigg)^{-1}}^{cd}F^{(1)}_{da}+
\f{1}{8}\nn&&{\bigg(M^{(0)}_+\bigg)^{-1}}^{ab}F^{(1)}_{ab}{\bigg(M^{(0)}_+
\bigg)^{-1}}^{cd}
F^{(1)}_{cd}\bigg]+\f{\mu_4}{2}\int \epsilon^{abcde} C_{a}F^{(1)}_{bc}F^{(1)}_{de},
\eea
where $M^{(0)}_+\equiv g+F^{(0)}$. From now on, we shall set $T_4=\mu_4$. For the above choice of the fields from   the NS-NS sector, RR sector and choosing the momentum as $k^{\mu}=(\omega,k_x,0,0)$, gives the same set of equation of motion for $A^{(1)}_t$ and  $A^{(1)}_x$ components as is found previously in eq(\ref{all_eom_gauge_fixed_5d}).
So, also the constraint equation. Denoting $E^{(1)}_x=\omega A^{(1)}_x+k_x A^{(1)}_t$, gives the same differential equation as before and also the expression of the $J^{(1)x}$ at a constant slice of $r$. It means the flow equation for the $\sigma^{xx}$ is same as before. Since, the solution to $A'_t$ is different means the flow equation for the conductivity in terms of charge density  becomes
\bea
\p_r \sigma^{xx}&=&\f{i\omega T_4\sqrt{g_{rr}}}{g_{xx}\sqrt{g_{tt}}[g_{xx}(g_{yy}g_{zz}+B^2)+(\rho-2B c^{(1)}_x)^2]^{3/2}}\Bigg[ \f{(\sigma^{xx})^2g^2_{xx}}{T^2_4}\Bigg(
(\rho-2B c^{(1)}_x)^2+\nn&&
\bigg[g_{xx}-\f{k^2_x}{\omega^2}g_{tt}\bigg]\bigg[g_{yy}g_{zz}+B^2\bigg]\Bigg)-\Bigg(
g_{xx}(g_{yy}g_{zz}+B^2)+(\rho-2B c^{(1)}_x)^2\Bigg)^2\Bigg].
\eea
Once again the regularity at the horizon requires us to set
\be
\label{bc_horizon_sigma_xx_dbi_cs}
\sigma^{xx}(r_h)=\Bigg[\f{T_4}{g_{xx}}\sqrt{g_{xx}(g_{yy}g_{zz}+B^2)+(\rho-2B c^{(1)}_x)^2}\Bigg]_{r_h},
\ee
which gives, at zero frequency, the dc conductivity at the boundary.

\subsubsection{The diffusion constant}
By following the previous arguments, in the hydrodynamic regime, $\omega \sim k^2_x\ra 0,~\omega \ll T$ and $k_x\ll T$. The flow equation can be approximated as
\be
\p_r \sigma^{xx}\simeq -i\f{k^2_x}{\omega}\f{(\sigma^{xx})^2 \sqrt{g_{rr}g_{tt}}~g_{xx}(g_{yy}g_{zz}+B^2)}{T_4[g_{xx}(g_{yy}g_{zz}+B^2)+(\rho-2B c^{(1)}_x)^2]^{3/2}},
\ee
which upon integrating from the horizon, $r=r_h$ to $r=\infty$, and using $G^{xx}_R(\omega,~k_x)=-i\omega \sigma^{xx}(r\ra\infty)=
\f{\omega^2\sigma^{xx}(r_h)}{i\omega-Dk^2_x}$, gives the diffusion constant as
\be
D=\f{\sigma^{xx}(r_h)}{T_4}\int^{\infty}_{r_h}
\f{\sqrt{g_{rr}g_{tt}}g_{xx}(g_{yy}g_{zz}+B^2)}{[(\rho-2B c^{(1)}_x)^2+g_{xx}(g_{yy}g_{zz}+B^2)]^{3/2}}.
\ee
\section{Conclusion}

In this paper, we have shown the existence of the linear temperature dependence of the resistivity for unit dynamical exponent in a different regime of the parameter space, namely, for strong magnetic field and low charge density limit by considering the DBI and CS type of action. In this setting: the  temperature(T),  the electric field and the magnetic field are introduced in the bulk and the charges are sitting outside the horizon, which sources the electric field. 
The motion of the charges in the presence of the electric and magnetic field generates non-trivial current both along the direction of the electric field  as well as in the perpendicular direction. Then Ohm's law suggests  the existence of the conductivity. 

With unit dynamical expoenent, the linear-$T$ behavior of the  resistivity,  which is the inverse of the conductivity, follows only along  the direction of the  applied magnetic field. Moreover, it does not show the necessary upturn behavior. However, the conductivity in the direction perpendicular to the magnetic field does show the upturn behavior but not the linear-T dependence of the resistivity.

From  fig(\ref{fig_3}) and fig(\ref{fig_4}), we do see that the resistivity along the direction of the magnetic field at zero temperature vanishes whereas
at finite temperature there exists a finite amount of resistivity.  The resistivity in the  direction  perpendicular to the magnetic field at zero temperature diverges but at finite temperature it becomes finite. This signals the possibility to see the metal-to-insulator transition.

There are still various physical quantities that are yet to be determined, satisfactorily. For example, with unit dynamical exponent, we can reproduce the desired  behavior of the resistivity in some corner of the parameter space, but in the same corner, it is not possible to find the desired behavior of the Hall angle. However, if we look at a different corner of  the parameter space then we can reproduce the Hall angle
very naturally but not the linear-T  behavior of the resistivity. With $z=2$, we do get the desired  result of the resistivity but not the Hall angle \cite{hpst} for NFL phase. So, it seems to find both the resistivity and the Hall angle of NFL state in a particular  setting is a daunting task for the application of the holographic  principle to condensed matter theories. 

There exists a universal behavior of the electrical conductivity  in a  specific regime of the parameter space, namely, in the large charge density and small magnetic field limit. From eq(\ref{bc_sigma_db_cs_2+1_d}) and  eq(\ref{bc_horizon_sigma_xx_dbi_cs}),  for $2+1$  and $3+1$ dimensional field theory, respectively,  with dynamical exponent $z$,  gives the longitudinal conductivity as $\sigma^{ii}\sim T^{-2/z}$ (no sum over $i$) and the Hall conductivity as $\sigma^{ij}\sim T^{-4/z}$, for $i\neq j$. The same result of the conductivity was obtained in  in any arbitrary spacetime dimension in that regime of the parameter space in \cite{ssp1}.  In this corner of the parameter space the Hall coefficient, $R_H$, goes as, $R_H\sim T^{4/z}$.

We have obtained the flow equation for both the electrical and magnetic conductivity.  At zero frequency the electric conductivity at the boundary is determined from its horizon value for trivial embeddings. However, for non-trivial embedding there arises some ambiguity in setting the regularity condition at the horizon, which we leave for future studies.\\

{\bf Acknowledgment:}\\
It is a pleasure to thank Ofer Aharony, Bum-Hoon Lee, Sudipta Mukherji and Sang-Jin Sin for useful discussions and encouragement.  This work is partly supported by the  Center for Quantum Spacetime (CQUeST), Seoul.


\begin{thebibliography}{99}
\bibitem{pmg} Paul. ~M.~Grant, ``The great quantum Conundrum," Nature, {\bf 476}, 37 (2011).
\bibitem{kj}K.~Jin, N.~P.~Butch, K.~Kirshenbaum, J.~Paglione and R.~L.~Greene, ``Link between spin fluctuations and electron pairing in copper oxide superconductors," Nature, {\bf 476}, 73 (2011).

\bibitem{daou} R.~Daou et al. ``Linear-T resistivity and change in Fermi surface at the pseudogap critical point of a high-$T_c$ superconductor," Nature, {\bf 5}, 31 (2009), [arXiv:0806.2881[cond-mat]].


\bibitem{jm}J. M. Maldacena,  ``The large N limit of superconformal field theories and supergravity,”
Adv. Theor. Math. Phys. {\bf 2}, 231 (1998), [arXiv:hep-th/9711200]; S. S. Gubser, I. R. Klebanov
and A. M. Polyakov,  ``Gauge theory correlators from noncritical string theory,”
Phys. Lett. {\bf B 428}, 105, (1998), [arXiv:hep-th/9802109]; E. Witten,  ``Anti-de Sitter
space and holography,” Adv. Theor. Math. Phys. {\bf 2}, 253 (1998), [arXiv:hep-th/9802150];
O. Aharony, S. S. Gubser, J. Maldacena H. Ooguri and Y. Oz,  ``Large N field theories,
String theory and gravity,” Phys. Rept, {\bf 323}, 183-386  (2000), [arXiv:hep-th/ 9905111].

\bibitem{kob}A. Karch and A. O'Bannon, ``Metallic AdS/CFT," JHEP {\bf 09}, 024 (2007),
[arXiv:0705.3870[hep-th]]

\bibitem{kk}A. Karch and E.~Katz, ``Adding flavors to AdS/CFT," JHEP {\bf 06}, 043 (2003), [arXiv:hep-th/0205236].



\bibitem{ob}A. O'Bannon, ``Hall Conductivity of Flavor Fields from
AdS/CFT," Phys. Rev. {\bf D 76}, 086007 (2007), [arXiv:0708.1994 [hep-th]].
\bibitem{hpst} S.~A.~Hartnoll, J.~Polchinski, E.~Silverstein and D.~Tong, ``Towards strange metallic holography,"  JHEP {\bf 1004}, 120 (2010), [arXiv:0912.1061[hep-th]].
\bibitem{kachru}K. Goldstein,  S. Kachru, S. Prakash, S. P. Trivedi and A. Westphal,
``Holography of Charged Dilaton Black Holes," JHEP {\bf 1008}, 078 (2010),
[arXiv:0911.3586[hep-th]].
\bibitem{kiritsis}
C. Charmousis, B. Goutraux, B. S. Kim, E. Kiritsis and R. Meyer,
``Effective Holographic
Theories for low-temperature condensed matter systems," JHEP {\bf 1011}, 151 (2010),   [arXiv:1005.4690[hepth]].


\bibitem{ks}
A. Karch and S. L. Sondhi, ``Non-linear, Finite Frequency, Quantum Critical Transport
from AdS/CFT," JHEP {\bf 1101}, 149 (2011), [arXiv:1008.4134[hep-th]].
\bibitem{ssp}
S. S. Pal, ``Model building in AdS/CMT: DC Conductivity and Hall angle," Phys. Rev. {\bf D 84}, 126009 (2011), 
[arXiv:1011.3117[hep-th]].


\bibitem{kkp}B. S. Kim, E. Kiritsis and C. Panagopoulos, ``Strange metal behavior from the Light-
Cone AdS Black Hole,"  [arXiv:1012.3646[hep-th]]
\bibitem{kim}K-Y. Kim, J. P. Shock and J. Tarrio, ``The open string membrane paradigm with external
electromagnetic fields," JHEP {\bf 1106}, 017 (2011), [arXiv:1103. 4581[hep-th]]; K-Y.
Kim and D-W. Pang, ``Holographic DC conductivities from the open string metric,"
[arXiv:1108.3791[hep-th]].
\bibitem{lpp}B-H. Lee, D-W. Pang and C. Park, ``A Holographic Model of Strange Metals," Int.
J. Mod. Phys. {\bf A 26}, 2279-2305 (2011), [arXiv:1107.5822[hep-th]].

\bibitem{il}N. Iqbal and H. Liu, ``Universality of the hydrodynamic limit in AdS/CFT and the
membrane paradigm," Phys. Rev. {\bf D 79}, 025023 (2009),  [arXiv:0809.3808[hep-th]].
\bibitem{ssp1} B-H.~Lee, S.~S.~Pal and S-J.~Sin, ``RG
flow of transport quantities," [arXiv:1108.5577[hep-th]].
\bibitem{akvk}J.~Alanen, E.~Keski-Vakkuri, P.~Kraus and V.~Suur-Uski, ``AC transport at Holographic Quantum Hall Transitions," JHEP {\bf 0911}, 014 (2009), [arXiv:0905.4538[hep-th]].

\bibitem{liu} T. ~Faulkner, H.~Liu, J.~McGreevy, and D.~Vegh, ``Emergent quantum criticality, Fermi surfaces and $AdS_2$," Phys. Rev. {\bf D 83}, 125002 (2011),[arXiv:0907.2694[hep-th]].
\bibitem{filmv} T. ~Faulkner, N.~ Iqbal, H.~Liu, J.~McGreevy, and D.~Vegh, ``Strange metal transport realized by gauge/gravity duality," Science, {\bf 329}, 1043 (2010); ``From black holes to strange metals," [arXiv:1003.1728[hep-th]].
\bibitem{fp} T. ~Faulkner and J.~Polchinski, ``Semi-Holographic Fermi liquid," JHEP {\bf 1106}, 012 (2011), [arXiv:1001.5049[hep-th]].
\bibitem{ss}S.~Sachdev, ``Holographic metals and the fractionalized Fermi liquid," Phys. Rev. Lett. {\bf 105}, 151602 (2010), [arXiv:1006.3794[hep-th]]. 


\bibitem{klm}S.~Kachru, X.~Liu and M.~Muligan, ``Gravity Duals of Lifshitz-like Fixed Points," Phys. Rev. {\bf D 78}, 106005 (2008), [arXiv:0808.1725[hep-th]].
\bibitem{pal}S. S. Pal, ``Anisotropic gravity solutions in AdS/CMT," [arXiv:0901.0599[hep-th]].

\bibitem{cm}K.~Copsey and R.~Mann, Pathologies in Asymptotically Lifshitz spacetimes," JHEP {\bf 1103}, 039 (2011), [arXiv:1011.3502[hep-th]].  

\bibitem{hw}G.~T.~Horowitz and B.~Way, ``Lifshitz Singularities," [arXiv:1111.1243[hep-th]].



\end{thebibliography}
\end{document}